\documentclass[journal=jctcce,manuscript=article]{achemso}

\usepackage{fontspec}
\usepackage{amsmath,amssymb,bm}
\usepackage{graphicx}
\usepackage{dcolumn}
\usepackage{booktabs}
\usepackage{xcolor}
\usepackage{hyperref}
\usepackage{orcidlink}
\usepackage{microtype}
\makeatletter
\providecommand{\@mcitecorrectmaxwidthsubitem}{0}
\makeatother

\hypersetup{
  colorlinks=true,
  linkcolor=blue,
  citecolor=blue,
  urlcolor=blue
}

\newcommand{\Tr}{\operatorname{Tr}}
\newcommand{\diag}{\operatorname{diag}}

\title{Quantum Boltzmann Equation for Static Correlation with Cumulant Kernel Closure}

\author{Romit Chakraborty\,\orcidlink{0000-0002-4638-6346}}
\affiliation{Point Reyes Sound, Inc., San Francisco, California 94114, USA}
\email{romit@pointreyessound.com}

\begin{document}

\begin{abstract}
    We present a Quantum Boltzmann Equation for static correlation by mapping the Bogoliubov-Born-Green-Kirkwood-Yvon hierarchy to discrete atomic-orbital matrix mechanics via the Wigner-Weyl isomorphism to evaluate scattering of the correlated two-body cumulant. This Quantum Boltzmann Equation converges to the active-space static correlation limit and preserves pure $N$-representability constraints during the kinetic relaxation. To resolve scaling bottlenecks of explicit two-body cumulant methods, we project the two-body cumulant onto its lowest unitary invariant and evaluate a susceptibility-windowed orbital scattering scale via a single Coulomb-exchange contraction. Relaxing onto a hyperbolic-tangent cumulant attractor through Markovian collisional relaxation yields a parameter-free Quantum Boltzmann Equation Cumulant-Kernel that regularizes derivative cusps without thermal smearing. Using the collinear $\mathrm{H}_3$ exchange reaction, the symmetric $D_{4h}$ square $\mathrm{H}_4$ dissociation, the non-adiabatic avoided crossing in $\mathrm{BeH}_2$, the symmetric bond cleavage of $\mathrm{H}_2\mathrm{O}$, the triple-bond dissociation of $\mathrm{N}_2$, the 11-molecule \texttt{MR-MGN-BE17} dissociation test set, and the singlet diradical isomers of benzyne as case studies, we show that this kinetic relaxation resolves strong static correlation at zero electronic temperature.
\end{abstract}

\section{Introduction}

Conventional electronic structure theory isolates the molecular ground state by solving a static matrix eigenvalue problem~\cite{szabo2012modern}. In contrast, phase-space formulations treat the electron cloud as a fluid governed by kinetic transport equations, allowing the evolution of probability density to be tracked in position and momentum space.~\cite{wigner1932quantum, weyl1927quantenmechanik} Early investigations by Dahl, Springborg, and others utilized the Wigner quasi-probability distribution to provide a phase-space view of density functional theory and atomic shell structures~\cite{dahl1982wigner, dahl2009moshinsky}, while recent advances by Subotnik and coworkers have established phase-space electronic Hamiltonians and Wigner-transformed many-body electronic friction for coupled nuclear--electronic dynamics~\cite{subotnik2016annurev, dou2017friction, dou2018perspective, tao2024phasespace, tao2025basisfree, bian2025phasespace}. In our previous work, we introduced the Quantum Boltzmann Equation Self-Consistent-Field (QBE-SCF)~\cite{chakraborty2026qbescf}. We demonstrated that relaxing the one-electron reduced density matrix via a Bhatnagar-Gross-Krook (BGK) collision operator~\cite{bhatnagar1954model} can entropically regularize mean-field singularities. By escaping the idempotency constraints of traditional solvers, QBE-SCF resolved the derivative discontinuities of the $\mathrm{H}_3$ dissociation and the $\mathrm{BeH}_2$ avoided crossing without invoking explicit multireference wavefunctions. To regularize diabatic traps, the single-particle system required the introduction of a finite electronic temperature to utilize macroscopic configurational entropy. To capture multi-reference electronic correlation at the zero-temperature limit without thermal broadening, we now extend phase-space collisional relaxation beyond the mean-field approximation.

In this work, we present a kinetic theory of multireference correlation and its cumulant-kernel closure. By coupling the one-particle distribution to the two-particle Wigner distribution via the first equation of the Bogoliubov-Born-Green-Kirkwood-Yvon (BBGKY) hierarchy~\cite{bogoliubov1946kinetic, born1946general, kirkwood1946statistical}, we derive the Quantum Boltzmann Equation for Static Correlation (Q-BESC) which captures both the single-particle mean-field energetics and the kinetics of multi-body momentum scattering under the Born-Oppenheimer approximation. By mapping the BBGKY hierarchy to discrete matrix mechanics, we replace thermal smearing with zero-temperature collisional relaxation governed by the correlated two-body cumulant (${}^{2}\boldsymbol{\Delta}$)~\cite{kutzelnigg1999cumulant, mazziotti1998contracted, mazziotti2006acse, mazziotti2007acse, mazziotti2012chemrev, mazziotti2016enhanced, smart2021cqe, liebert2025spin}. We demonstrate that this two-body collision operator resolves strong zero-temperature static correlation, and converges to the Complete Active Space ($\mathrm{CASCI}$/$\mathrm{CASSCF}$) and Density Matrix Renormalization Group (DMRG)~\cite{white1992density, chan2002highly} static correlation limits which coincide with Full Configuration Interaction, FCI, in few-electron systems. To overcome the combinatorial cost of explicit two-body tensors, we project this collision operator onto the lowest unitary invariant of the cumulant, closing the hierarchy within the one-electron density matrix to formulate the Quantum Boltzmann Equation Cumulant-Kernel (Q-BECK). Through case studies in static correlation spanning the $\mathrm{H}_3$ collinear exchange reaction, the square $\mathrm{H}_4$ ($D_{4h}$) dissociation, the $\mathrm{BeH}_2$ avoided crossing, the symmetric double dissociation of $\mathrm{H}_2\mathrm{O}$, the $\mathrm{N}_2$ triple-bond cleavage, the \textit{ortho}-, \textit{meta}-, and \textit{para}-benzyne biradical isomers, and the 11-molecule \texttt{MR-MGN-BE17} dissociation test set at $5.0\,\text{\AA}$, we show that Q-BECK delivers smooth, cusp-free potential energy surfaces and accurate biradical splittings. Finally, we analyze the scaling and memory requirements of Q-BECK to find quadratic one-body storage costs and near-cubic empirical time scaling in dense regimes, which reduces toward near-quadratic scaling in spatially extended systems via Schwarz screening of localized atomic orbitals.

\section{Mapping of the Many-Body BBGKY Hierarchy}

To model two-body electronic correlations in phase space, we project the joint two-particle Wigner quasi-probability distribution $W^{(2)}(r_1, p_1, r_2, p_2, t)$ onto a stationary atomic-orbital (AO) basis $\{\chi_{\mu}\}$ under the Born--Oppenheimer approximation:
\begin{equation}
W^{(2)}(r_1, p_1, r_2, p_2, t) = \sum_{\mu\nu\lambda\sigma} {}^{2}\mathbf{D}_{\mu\nu\lambda\sigma}(t)\,\mathcal{W}_{\mu\nu\lambda\sigma}^{(2)}(r_1, p_1, r_2, p_2),
\end{equation}
where all temporal evolution is isolated within the two-particle reduced density matrix (2-RDM) ${}^{2}\mathbf{D}_{\mu\nu\lambda\sigma}(t)$ and the static two-particle phase-space basis tetrad is
\begin{equation}
\begin{split}
\mathcal{W}_{\mu\nu\lambda\sigma}^{(2)}(r_1, p_1, r_2, p_2) &= \frac{1}{(2\pi)^6} \iint d^3s_1 \, d^3s_2 \, e^{-i(p_1\cdot s_1 + p_2\cdot s_2)} \\
&\quad \times \chi_{\mu}\bigl(r_1 + \tfrac{s_1}{2}\bigr) \chi_{\nu}^{*}\bigl(r_1 - \tfrac{s_1}{2}\bigr) \chi_{\lambda}\bigl(r_2 + \tfrac{s_2}{2}\bigr) \chi_{\sigma}^{*}\bigl(r_2 - \tfrac{s_2}{2}\bigr).
\end{split}
\end{equation}
In the first equation of the BBGKY hierarchy, the single-particle Wigner function $W^{(1)}$ couples to $W^{(2)}$ through the inter-electronic Coulomb interaction ${}^{2}\mathbf{V}_{ee}$:
\begin{equation}
\frac{\partial W^{(1)}}{\partial t} + \frac{p_1}{m} \cdot \nabla_{r_1} W^{(1)} + \mathcal{F}_{\mathrm{ext}} \cdot \nabla_{p_1} W^{(1)} = \mathcal{I}_{\mathrm{coll}}\bigl[W^{(2)}, {}^{2}\mathbf{V}_{ee}\bigr],
\end{equation}
where the quantum collision integral $\mathcal{I}_{\mathrm{coll}}$ scatters correlated pair distributions rather than independent particles in a static mean field. Invoking the Wigner--Weyl isomorphism~\cite{wigner1932quantum, weyl1927quantenmechanik, dou2017friction, dou2018perspective, tao2024phasespace, tao2025basisfree, bian2025phasespace} maps the continuous phase-space Moyal bracket onto the Hilbert-space commutator, transforming the first BBGKY equation into the exact contracted Liouville equation for the 1-RDM (${}^{1}\mathbf{P}$):
\begin{equation}
\label{eq:bbgky_commutator}
\frac{\partial {}^{1}\mathbf{P}}{\partial t} + i\bigl[{}^{1}\mathbf{H}_{\mathrm{core}}, {}^{1}\mathbf{P}\bigr] = -i \Tr_2 \bigl[{}^{2}\mathbf{V}_{ee}, {}^{2}\mathbf{D}\bigr],
\end{equation}
where ${}^{1}\mathbf{H}_{\mathrm{core}}$ is the one-electron core Hamiltonian and $\Tr_2$ denotes the partial trace over the second particle coordinate.

\section{Pair-Correlation Decomposition and Cumulant Closure}

To incorporate multi-body correlations without an external wavefunction oracle, we decompose the 2-RDM into its antisymmetrized mean-field wedge product and the correlated two-body cumulant ${}^{2}\boldsymbol{\Delta}$, partitioned across three orthogonal interaction channels~\cite{kutzelnigg1999cumulant, mazziotti1998contracted, mazziotti2012chemrev, liebert2025spin}:
\begin{equation}
\label{eq:cumulant_decomp}
{}^{2}\mathbf{D} = {}^{1}\mathbf{P} \wedge {}^{1}\mathbf{P} + {}^{2}\boldsymbol{\Delta}[{}^{1}\mathbf{P}], \qquad {}^{2}\boldsymbol{\Delta}[{}^{1}\mathbf{P}] = {}^{2}\boldsymbol{\Delta}_0[{}^{1}\mathbf{P}] + {}^{2}\boldsymbol{\Delta}_1[{}^{1}\mathbf{P}] + {}^{2}\boldsymbol{\Delta}_2[{}^{1}\mathbf{P}],
\end{equation}
which encode the Borland--Dennis bounds of the generalized Pauli exclusion principle~\cite{borland1972conditions, klyachko2006quantum, schilling2013pinning, chakraborty2014generalized, chakraborty2015structure, chakraborty2017thesis, chakraborty2018sparsity, schilling2019diverging, schilling2020implications1, maciazek2020implications2, schilling2021ensemble, liebert2022foundation, liebert2023refining, liebert2025toolbox, liebert2025ensemble, liebert2025partial} and active--virtual orbital entrainment. Absorbing the mean-field contraction $\Tr_2[{}^{2}\mathbf{V}_{ee}, {}^{1}\mathbf{P} \wedge {}^{1}\mathbf{P}] = [{}^{1}\mathbf{G}[{}^{1}\mathbf{P}], {}^{1}\mathbf{P}]$ into the effective Fock operator ${}^{1}\mathbf{F}[{}^{1}\mathbf{P}] = {}^{1}\mathbf{H}_{\mathrm{core}} + {}^{1}\mathbf{G}[{}^{1}\mathbf{P}]$ isolates the correlated two-body collision commutator $\mathcal{Q}_{\mathrm{corr}}[{}^{1}\mathbf{P}] = -i \Tr_2[{}^{2}\mathbf{V}_{ee}, {}^{2}\boldsymbol{\Delta}[{}^{1}\mathbf{P}]]$ in the contracted transport equation:
\begin{equation}
\label{eq:closed_liouville}
\frac{\partial {}^{1}\mathbf{P}}{\partial t} + i\bigl[{}^{1}\mathbf{F}[{}^{1}\mathbf{P}], {}^{1}\mathbf{P}\bigr] = -i \Tr_2 \bigl[{}^{2}\mathbf{V}_{ee}, {}^{2}\boldsymbol{\Delta}[{}^{1}\mathbf{P}]\bigr] \equiv \mathcal{Q}_{\mathrm{corr}}[{}^{1}\mathbf{P}],
\end{equation}
whose stationary ground-state relaxation is driven by modeling the dissipative correlation drift through the functional derivative $\delta E_{\mathrm{corr}}[{}^{1}\mathbf{P}] / \delta {}^{1}\mathbf{P}$ of the static correlation energy $E_{\mathrm{corr}}[{}^{1}\mathbf{P}]$.

In the full Quantum Boltzmann Equation for Static Correlation (Q-BESC), all three cumulant channels in Eq.~\eqref{eq:cumulant_decomp}—including the traceless one-body contraction ${}^{2}\boldsymbol{\Delta}_1$ and the contractionless two-body residual ${}^{2}\boldsymbol{\Delta}_2$—are retained within a correlated $(N_{\mathrm{act}}, M_{\mathrm{act}})$ valence active manifold $\mathcal{A}$ (spanning the full orbital space in few-electron benchmarks) while core and virtual subspaces remain mean-field separable (${}^{2}\boldsymbol{\Delta}\big|_{\mathcal{A}^\perp} = \mathbf{0}$). Projecting the active-space cumulant tensor ${}^{2}\boldsymbol{\Delta}_{\mathcal{A}} = {}^{2}\boldsymbol{\Delta}_0 + {}^{2}\boldsymbol{\Delta}_1 + {}^{2}\boldsymbol{\Delta}_2$ into the AO phase-space representation closes the collision integral in Eq.~\eqref{eq:closed_liouville} via contracted two-body Markovian relaxation coupled to active--inactive orbital stationarity, $[{}^{1}\mathbf{F}_{\mathrm{stat}}, {}^{1}\mathbf{P}_{\mathrm{stat}}] = \mathbf{0}$, where the stationary generalized Fock generator absorbs both the mean-field field ${}^{1}\mathbf{F}[{}^{1}\mathbf{P}_{\mathrm{stat}}]$ and the contracted cumulant scattering $\Tr_2[{}^{2}\mathbf{V}_{ee}, {}^{2}\boldsymbol{\Delta}_{\mathcal{A}}]$. At kinetic equilibrium ($\partial_t {}^{1}\mathbf{P} = \mathbf{0}$), this coupled relaxation satisfies the contracted one-body Brillouin stationarity condition and evaluates the total electronic energy $E_{\mathrm{Q\text{-}BESC}} = \Tr({}^{1}\mathbf{H}_{\mathrm{core}}\,{}^{1}\mathbf{P}_{\mathrm{stat}}) + \frac{1}{2}\Tr\bigl({}^{2}\mathbf{V}_{ee}({}^{1}\mathbf{P}_{\mathrm{stat}} \wedge {}^{1}\mathbf{P}_{\mathrm{stat}} + {}^{2}\boldsymbol{\Delta}_{\mathcal{A}})\bigr) + E_{\mathrm{nuc}}$, recovering the exact active-space static correlation limit while preserving pure-state $N$-representability along the relaxation trajectory.

\subsection{Unitary Invariant and Q-BECK Discretization}

Following the unitary group $U(r)$ decomposition of Gibney, Boyn, and Mazziotti~\cite{gibney2021sdp, gibney2022rdmft, boyn2022orbital, gibney2023universal, liebert2025spin} in a manifold of $r$ spin-orbitals, the channels in Eq.~\eqref{eq:cumulant_decomp} correspond to the fully symmetric trace component ${}^{2}\boldsymbol{\Delta}_0$, the traceless one-body contraction ${}^{2}\boldsymbol{\Delta}_1$, and the traceless, contractionless residual ${}^{2}\boldsymbol{\Delta}_2$. Projecting ${}^{2}\boldsymbol{\Delta}$ onto its lowest-order unitary invariant,
\begin{equation}
{}^{2}\boldsymbol{\Delta}_0 = \frac{2}{r(r-1)}\Tr({}^{2}\boldsymbol{\Delta})\,{}^{2}\mathbf{I}, \qquad \Tr({}^{2}\boldsymbol{\Delta}) = -\frac{1}{2}\Tr\bigl({}^{1}\mathbf{P} - {}^{1}\mathbf{P}^2\bigr),
\end{equation}
where $({}^{2}\mathbf{I})^{ij}_{kl} = \frac{1}{2}(\delta^i_k \delta^j_l - \delta^i_l \delta^j_k)$ satisfies $\Tr({}^{2}\mathbf{I}) = \binom{r}{2}$, isolates the lowest-order static correlation functional and its closed $\mathcal{O}(N_{\mathrm{AO}}^3)$ correlation drift derivative~\cite{gibney2021sdp, gibney2022rdmft, gibney2023universal}:
\begin{equation}
\label{eq:e0_and_drift}
E_0^{\Delta}[{}^{1}\mathbf{P}] = \Tr\bigl({}^{2}\mathbf{V}_{ee}\,{}^{2}\boldsymbol{\Delta}_0\bigr) = -\sum_{i}\omega_{i}\,\bigl({}^{1}\mathbf{P} - {}^{1}\mathbf{P}^2\bigr)_{ii}, \qquad \frac{\delta E_0^{\Delta}[{}^{1}\mathbf{P}]}{\delta {}^{1}\mathbf{P}} = -\diag(\boldsymbol{\omega})\bigl(\mathbf{I} - 2\,{}^{1}\mathbf{P}\bigr).
\end{equation}
The size-consistent orbital scattering scale $\omega_i$ modulates the susceptibility-weighted antisymmetrized two-electron integrals over spatial orbitals $\{\tilde{i},\tilde{j}\}$ by a smooth, symmetric valence-band susceptibility envelope $w_i \in [0, 1]$ centered at the frontier midpoint $\mu_0 = \frac{1}{2}(\epsilon_{\mathrm{HOMO}} + \epsilon_{\mathrm{LUMO}})$, screening out deep core ($\epsilon_j \ll \mu_0$) and high virtual ($\epsilon_j \gg \mu_0$) orbitals without user-defined active manifolds $\mathcal{A}$. Subtracting the diagonal $j = i$ self-interaction integral $g_{ii} = \langle\tilde{i}\tilde{i}|\tilde{i}\tilde{i}\rangle$, $\omega_i$ is evaluated via a single $\mathcal{O}(N_{\mathrm{AO}}^3)$ AO contraction $\mathbf{G}[\mathbf{P}_w] = 2\mathbf{J}[\mathbf{P}_w] - \mathbf{K}[\mathbf{P}_w]$ on the weighted density $\mathbf{P}_w = \sum_j w_j \mathbf{c}_j \mathbf{c}_j^{\mathsf{T}}$:
\begin{equation}
\label{eq:omega_scale}
\omega_{i} = w_i\,\frac{\sum_{j \neq i} w_j \bigl(2\langle\tilde{i}\tilde{j}|\tilde{i}\tilde{j}\rangle - \langle\tilde{i}\tilde{j}|\tilde{j}\tilde{i}\rangle\bigr)}{\sum_{j \neq i} w_j} = w_i\,\frac{\mathbf{c}_i^{\mathsf{T}}\mathbf{G}[\mathbf{P}_w]\mathbf{c}_i - w_i g_{ii}}{\sum_{j \neq i} w_j},
\end{equation}
where the susceptibility window width $\Delta_0 \sim \mathcal{O}(0.5\,E_{\mathrm{h}})$ is fixed universally at the atomic-unit valence binding scale (separating covalent valence shells from deep core states $\epsilon_j \ll \mu_0$ and high-energy virtual continua $\epsilon_j \gg \mu_0$) across all molecules, spin states, basis sets, and nuclear geometries without empirical fitting or per-system tuning, and diagonal self-interaction subtraction and numerical floor regularization are handled internally within the $\mathcal{O}(N_{\mathrm{AO}}^3)$ Q-BECK kernel. Because $\Delta_0$ enters only the self-normalized core/virtual partition envelope $w_j / \sum_{j \neq i} w_j$ while the correlation amplitude is set by the ab initio two-electron expectation scale $\omega_i$, the resulting closure introduces zero adjustable correlation parameters.

Combining this cumulant correlation drift with BGK mean-field relaxation~\cite{chakraborty2026qbescf}, dropping the conservative Liouville streaming commutator to isolate the stationary ground state ($[\mathbf{F}_{\mathrm{stat}}, \mathbf{P}_{\mathrm{stat}}] = 0$), and discretizing over $\Delta t$ with Markovian relaxation weight $w = \Delta t/\tau$ yields the $\mathcal{O}(N^3)$ Quantum Boltzmann Equation Cumulant-Kernel (Q-BECK) update:
\begin{equation}
\label{eq:qbeck_update}
{}^{1}\mathbf{P}(t+\Delta t) = (1 - w){}^{1}\mathbf{P}(t) + w\,{}^{1}\mathbf{P}^{\mathrm{eq}}\bigl[{}^{1}\mathbf{F}(t)\bigr] - \Delta t\,\diag(\boldsymbol{\omega})\bigl(\mathbf{I} - 2\,{}^{1}\mathbf{P}(t)\bigr).
\end{equation}
Folding the correlation drift into the self-consistent effective Fock generator,
\begin{equation}
{}^{1}\mathbf{F}_{\mathrm{eff}}[{}^{1}\mathbf{P}] = {}^{1}\mathbf{F}[{}^{1}\mathbf{P}] - \diag(\boldsymbol{\omega})\bigl(\mathbf{I} - 2\,{}^{1}\mathbf{P}\bigr),
\end{equation}
and enforcing particle-number conservation $\Tr({}^{1}\mathbf{P}) = N_{\mathrm{elec}}$ via a Lagrange chemical potential $\mu$ produces the analytical $C^\infty$ hyperbolic-tangent cumulant attractor:
\begin{equation}
\label{eq:qbeck_attractor}
n_i^{\mathrm{eq}} = \frac{1}{2}\left[1 - \tanh\!\left(\frac{\epsilon_i - \mu}{2\omega_{i}}\right)\right],
\end{equation}
guaranteeing ensemble Pauli $N$-representability ($0 \le n_i^{\mathrm{eq}} \le 1$) and smooth differentiability across all nuclear geometries at $T_{\mathrm{elec}} = 0$.

\subsection{Computational Methodology}
\label{sec:computational_methodology}

All QBE-SCF, Q-BECK, and Q-BESC calculations are executed in AO matrix mechanics using one- and two-electron integrals from PySCF~\cite{sun2018pyscf} under six algorithmic specifications:
\begin{enumerate}
    \item \textbf{Markovian Relaxation ($w = \Delta t / \tau = 0.5$)}: At each micro-step $k$, the AO 1-RDM updates via ${}^{1}\mathbf{P}^{(k+1)} = (1 - w){}^{1}\mathbf{P}^{(k)} + w\,{}^{1}\mathbf{P}^{\mathrm{eq}}[{}^{1}\mathbf{F}_{\mathrm{eff}}({}^{1}\mathbf{P}^{(k)})]$ with $w = 0.5$, preserving the Pauli spectrum $[0, 1]$ without DIIS history buffers or orbital-Hessian inversions.
    \item \textbf{Stationarity Thresholds ($|\Delta E| < 10^{-9}\,E_h$, $\Vert[\mathbf{F}_{\mathrm{eff}}, {}^{1}\mathbf{P}]\Vert < 10^{-9}$)}: Convergence is certified when $|\Delta E^{(k)}| < 10^{-9}\,E_h$, $\Vert{}^{1}\mathbf{P}^{(k+1)} - {}^{1}\mathbf{P}^{(k)}\Vert_{\mathrm{F}} < 10^{-9}$, and the Löwdin-orthogonalized Liouville commutator satisfies
    \begin{equation}
        \bigl\Vert \bigl[\mathbf{S}^{-1/2}\,{}^{1}\mathbf{F}_{\mathrm{eff}}\,\mathbf{S}^{-1/2},\; \mathbf{S}^{1/2}\,{}^{1}\mathbf{P}\,\mathbf{S}^{1/2}\bigr] \bigr\Vert_{\mathrm{F}} < 10^{-9}.
    \end{equation}
    \item \textbf{Chemical-Potential Bracketing ($10^{-12}$)}: At every micro-step, $\mu$ in Eq.~\eqref{eq:qbeck_attractor} is solved via root-finding on $\sum_{i=1}^{N_{\mathrm{AO}}} n_i^{\mathrm{eq}}(\mu) - N_{\mathrm{elec}} = 0$ to $\delta\mu \le 10^{-12}\,E_{\mathrm{h}}$, enforcing $\Tr({}^{1}\mathbf{P}\,\mathbf{S}) = N_{\mathrm{elec}}$ to machine precision.
    \item \textbf{Virial Diagnostic and Natural Occupations ($n_i^{\mathrm{spatial}}$)}: Equilibrium is monitored via the virial ratio $-\langle V\rangle / 2\langle T\rangle = -(E_{\mathrm{tot}} - \Tr({}^{1}\mathbf{P}\mathbf{T})) / (2\Tr({}^{1}\mathbf{P}\mathbf{T}))$ and spatial eigenvalues $n_i^{\mathrm{spatial}} \in [0, 2]$ of $\mathbf{P}_{\mathrm{tot}}\mathbf{S}$, verifying that core/spectator orbitals remain idempotent ($n_i^{\mathrm{spatial}} \in \{0, 2\}$) while bond-breaking orbitals approach the Mott limit ($n_i^{\mathrm{spatial}} \to 1$).
    \item \textbf{Zero-Temperature Replacement of Thermal Smearing ($T_{\mathrm{elec}} = 0.06\,\mathrm{Ha} \to \omega_i$)}: Unlike Paper~1~\cite{chakraborty2026qbescf}, which required an empirical Fermi--Dirac temperature $T_{\mathrm{elec}} = 0.06\,\mathrm{Ha}$ that uniformly broadened all orbitals, Q-BECK operates at $T_{\mathrm{elec}} = 0$ by replacing $k_{\mathrm{B}}T_{\mathrm{elec}}$ with the orbital-resolved scattering rate $\omega_i$ evaluated from $\mathbf{G}[\mathbf{P}_w]$.
    \item \textbf{Spin-Symmetry Sector Selection (${}^{1}\mathbf{P}_\alpha = {}^{1}\mathbf{P}_\beta$ vs.\ ${}^{1}\mathbf{P}_\alpha \neq {}^{1}\mathbf{P}_\beta$)}: Spin-restricted singlet manifolds (${}^{1}\mathbf{P}_\alpha = {}^{1}\mathbf{P}_\beta = \frac{1}{2}\,{}^{1}\mathbf{P}$, $\langle\hat{S}^2\rangle = 0$) are enforced whenever isolating pure singlet states against lower-lying triplet contamination—specifically across the $\mathrm{BeH}_2(S_0)$ avoided crossing, the \textit{ortho}-, \textit{meta}-, and \textit{para}-benzyne biradicals, the equilibrium $\mathrm{N}_2$ basis hierarchy (Table~\ref{tab:n2_eq_convergence}), and the \texttt{MR-MGN-BE17} static-correlation benchmark (Block~I of Table~\ref{tab:mr_mgn_be17})—whereas spin-unrestricted AO relaxation (${}^{1}\mathbf{P}_\alpha \neq {}^{1}\mathbf{P}_\beta$) is permitted along open-shell homolytic dissociation coordinates ($\mathrm{H}_3$, $\mathrm{H}_4$, $\mathrm{H}_2\mathrm{O}$, and stretched $\mathrm{N}_2$). Along $\mathrm{N}_2$ dissociation, the crossover from delocalized singlet cumulant fractionalization to broken-symmetry spin localization across $R \in [1.7, 1.9]\,\text{\AA}$ directly produces the intermediate shoulder in Figure~\ref{fig:n2_dissociation}a and the modest asymptotic offset beneath $\mathrm{Q\text{-}BESC}(6,6)$ beyond $R \approx 2.2\,\text{\AA}$ via unrestricted exchange spin-polarization of the $2s$ sub-valence shells on the high-spin ($S_A = 3/2$) nitrogen atoms.
\end{enumerate}

\section{Case Studies in Static Correlation}

To verify that explicit two-body cumulant scattering (${}^{2}\boldsymbol{\Delta}$) in Q-BESC (\textit{Quantum Boltzmann Equation for Static Correlation}) and its cumulant-kernel projection in Q-BECK replace artificial thermal ergodicity with physical zero-temperature static correlation at $T_{\mathrm{elec}} = 0$, we evaluate a comprehensive benchmarking suite spanning single- and multi-bond dissociations, radical reaction pathways, non-adiabatic avoided crossings, and strongly entangled multireference states. Across Figures~\ref{fig:h3_gpc_analysis}--\ref{fig:benzyne_benchmark}, the $\mathcal{O}(N^3)$ Q-BECK trajectory is plotted as a high-contrast gold curve alongside the uncorrelated 1-particle QBE-SCF baseline and the exact-cumulant Q-BESC solver. Crucially, while Q-BESC converges to the exact $\mathrm{CASCI}$/$\mathrm{CASSCF}$ active-space static correlation limit—which coincides with full-basis FCI in few-electron systems such as $\mathrm{H}_3$ and $\mathrm{H}_4$—any uniform equilibrium energy offset relative to single-reference dynamical correlation methods ($\mathrm{MP2}$/$\mathrm{CCSD}$ in $\mathrm{H}_2\mathrm{O}$) or full experimental dissociation energies ($D_e$ in $\mathrm{N}_2$) reflects out-of-space dynamic correlation outside the valence active space.

\subsection{Collinear \texorpdfstring{$\mathrm{H}_3$}{H3} Exchange Reaction and Generalized Pauli Polytope Distances}
\label{sec:h3_reaction_coordinate}

\begin{figure}[!t]
    \centering
    \includegraphics[width=0.98\linewidth]{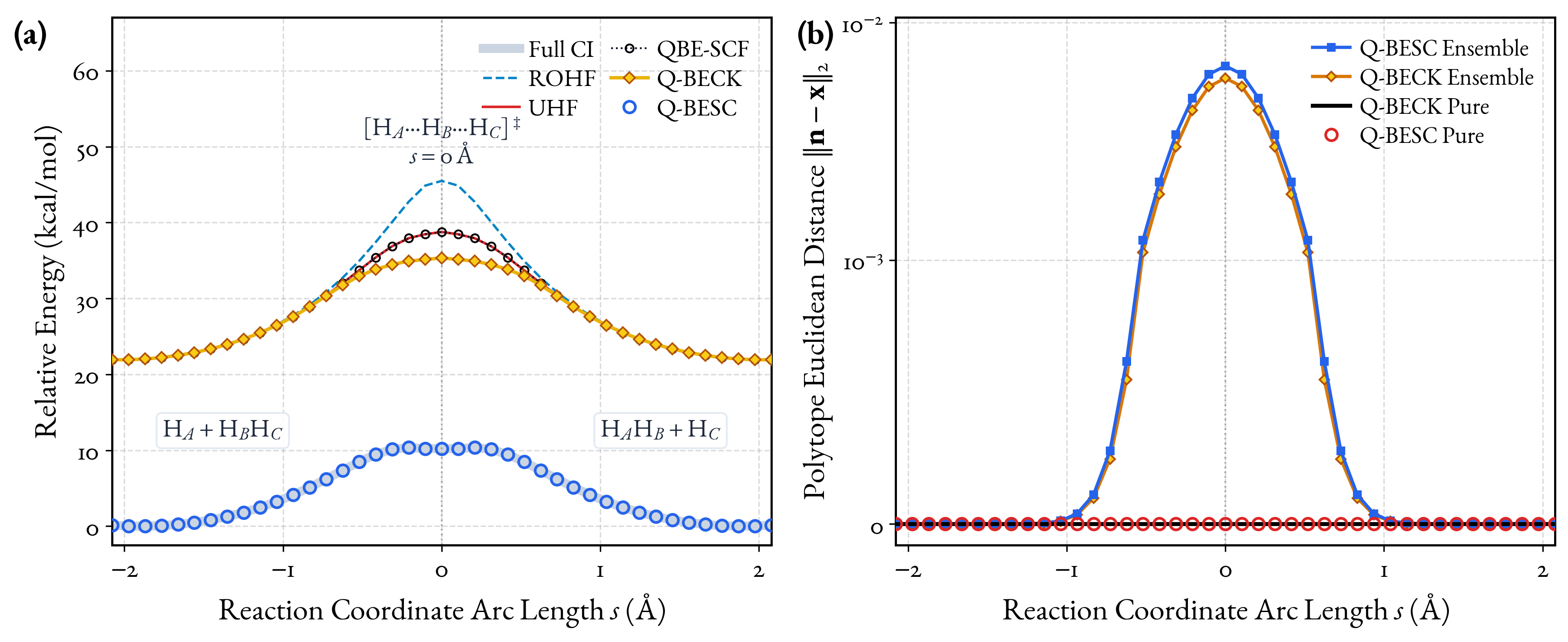}
    \caption[Potential energy barrier and generalized Pauli polytope distances for the collinear H3 exchange reaction]{\textbf{Collinear $\mathrm{H}_3$ minimum-energy-path (MEP) reaction barrier (a) and constrained minimum Euclidean polytope distances (b) across $s \in [-2.08, +2.08]\,\text{\AA}$ in the \texttt{cc-pVDZ} basis.}
    In Panel~(a), Q-BESC reproduces the exact Full CI activation barrier of $10.22\,\mathrm{kcal/mol}$ at the symmetric transition state ($s = 0\,\text{\AA}$) to sub-nanohartree precision, whereas ROHF ($23.46\,\mathrm{kcal/mol}$) and unrestricted UHF/QBE-SCF ($16.74\,\mathrm{kcal/mol}$) substantially overestimate the barrier height.
    Relaxing the $\mathcal{O}(N^3)$ Q-BECK solver onto its smooth hyperbolic-tangent cumulant attractor lowers the activation barrier to $12.76\,\mathrm{kcal/mol}$, recovering over $60\%$ of the transition-state static correlation energy without derivative discontinuities.
    In Panel~(b), the constrained minimum Euclidean distances~\cite{chakraborty2014generalized} of the $\mathrm{CAS}(3,3)$ spin-orbital occupation spectrum $\bm{n}$ to the pure Borland--Dennis facet ($\|\bm{n} - \bm{p}\|_2 = 0$, rendered as open red circles for Q-BESC and a solid black line for Q-BECK) and the ensemble Pauli boundary ($\|\bm{n} - \bm{e}\|_2$) confirm that both Q-BESC and Q-BECK remain pinned to the pure facet while forming symmetric bell-shaped ensemble-distance peaks centered at $s = 0\,\text{\AA}$.}
    \label{fig:h3_gpc_analysis}
    \label{fig:h3_reaction_coordinate}
\end{figure}

To evaluate non-equilibrium bond rearrangement and pure-state $N$-representability along a reactive coordinate without a parent wavefunction, we examine the collinear hydrogen exchange reaction $\mathrm{H}_A + \mathrm{H}_B\mathrm{H}_C \to [\mathrm{H}_A \cdots \mathrm{H}_B \cdots \mathrm{H}_C]^\ddagger \to \mathrm{H}_A\mathrm{H}_B + \mathrm{H}_C$ in the \texttt{cc-pVDZ} basis ($15$ spatial orbitals, $N=3$, $S=1/2$) across the minimum-energy-path (MEP) arc length $s \in [-2.08, +2.08]\,\text{\AA}$, connecting the reactant/product channels ($s = \pm 2.08\,\text{\AA}$, $\{R_1, R_2\} = \{3.00, 0.74\}\,\text{\AA}$) through the symmetric transition state ($s = 0\,\text{\AA}$, $R_1 = R_2 = 0.93\,\text{\AA}$). Beyond the standard ensemble Pauli bounds ($0 \le n_i \le 1$), pure $N$-electron wavefunctions restrict the ordered natural spin-orbital occupation spectrum $\bm{n} = (n_1 \ge \dots \ge n_6)$ ($\sum_{i=1}^6 n_i = 3$) of the $\mathrm{CAS}(3,3)$ valence manifold to a convex pure-state polytope $\mathcal{P}_1^3 \subset \mathcal{E}_1^3$ defined by the generalized Pauli exclusion principle~\cite{borland1972conditions, klyachko2006quantum, schilling2013pinning, chakraborty2014generalized, chakraborty2015sufficient, chakraborty2015structure, chakraborty2017thesis, chakraborty2018sparsity, schilling2018generalized, schilling2019diverging, schilling2020implications1, maciazek2020implications2, benavides2020bosons, schilling2021ensemble, liebert2022foundation, liebert2023bosonic, liebert2023excited, liebert2023refining, liebert2025toolbox, liebert2025ensemble, liebert2025partial}, governed by the Borland--Dennis equalities ($n_1 + n_6 = n_2 + n_5 = n_3 + n_4 = 1$) and polygonal slack inequality $D_{\mathrm{BD}} \equiv n_5 + n_6 - n_4 \ge 0$~\cite{borland1972conditions}. Following Chakraborty and Mazziotti~\cite{chakraborty2014generalized, chakraborty2015sufficient, chakraborty2015structure, chakraborty2017thesis, chakraborty2018sparsity, liebert2025spin, liebert2025partial} and Schilling and coworkers~\cite{schilling2013pinning, schilling2018generalized, schilling2019diverging, schilling2020implications1, maciazek2020implications2, schilling2021ensemble, liebert2022foundation, liebert2023bosonic, liebert2023excited, liebert2023refining, liebert2025toolbox, liebert2025ensemble}, we track the geometric evolution of the kinetically relaxed 1-RDM along $s$ using the constrained minimum Euclidean ($l_2$) distances to the ensemble Pauli hypercube boundary $\partial\mathcal{E}_1^N$, $\|\bm{n} - \bm{e}\|_2 = \sqrt{6/5}\,\min(1 - n_1, n_6)$, and to the pure Borland--Dennis facet $\partial\mathcal{P}_1^N$ ($p_1 + p_6 = p_2 + p_5 = p_3 + p_4 = 1$, $p_5 + p_6 - p_4 = 0$), $\|\bm{n} - \bm{p}\|_2 = \min_{\bm{p} \in \partial\mathcal{P}_1^N} \|\bm{n} - \bm{p}\|_2$.

As shown in Figure~\ref{fig:h3_reaction_coordinate}a, exact-cumulant Q-BESC reproduces the Full CI potential energy profile ($E_{\mathrm{FCI}} = E_{\mathrm{Q\text{-}BESC}} = -1.646515\,E_h$ at $s = 0\,\text{\AA}$, $\Delta E^\ddagger = 16.29\,\mathrm{m}E_h = 10.22\,\mathrm{kcal/mol}$) across all 41 MEP points to sub-nanohartree precision ($\max_s |E_{\mathrm{Q\text{-}BESC}} - E_{\mathrm{FCI}}| = 2.91 \times 10^{-10}\,E_h$). While ROHF ($\Delta E^\ddagger = 23.46\,\mathrm{kcal/mol}$) and single-particle UHF/QBE-SCF ($T_{\mathrm{elec}} = 0$, $E_{\mathrm{QBE\text{-}SCF}} = -1.601033\,E_h$, $|E_{\mathrm{QBE\text{-}SCF}} - E_{\mathrm{UHF}}| < 10^{-9}\,E_h$, $\Delta E^\ddagger = 16.74\,\mathrm{kcal/mol}$) substantially overestimate the barrier height, relaxing the $\mathcal{O}(N^3)$ Q-BECK scheme onto its smooth $C^\infty$ hyperbolic-tangent cumulant attractor ($E_{\mathrm{Q\text{-}BECK}} = -1.607527\,E_h$ at $s = 0\,\text{\AA}$, virial ratio $-\langle V\rangle/2\langle T\rangle = 1.009775$, $S_{\mathrm{config}} = 1.279\,\mathrm{bits}$, $\Vert[\mathbf{F}_{\mathrm{eff}}, {}^{1}\mathbf{P}]\Vert_{\mathrm{F}} = 8.17 \times 10^{-10}$ in $65$ cycles) yields a smooth bell-shaped barrier of $\Delta E_{\mathrm{Q\text{-}BECK}}^\ddagger = 12.76\,\mathrm{kcal/mol}$ ($20.34\,\mathrm{m}E_h$), recovering over $60\%$ of the transition-state static correlation energy. Simultaneously, Figure~\ref{fig:h3_reaction_coordinate}b demonstrates that both Q-BESC (open red circles) and Q-BECK (solid black line) remain pinned to the pure Borland--Dennis facet ($\|\bm{n} - \bm{p}\|_2 = 0$, $D_{\mathrm{BD}} \equiv n_5 + n_6 - n_4 \ge 0$) throughout the reaction while detaching from the ensemble boundary $\partial\mathcal{E}_1^N$ ($0 = \|\bm{n} - \bm{p}\|_2 < \|\bm{n} - \bm{e}\|_2$)~\cite{chakraborty2014generalized, chakraborty2015structure, schilling2013pinning, schilling2019diverging, schilling2020implications1, liebert2025toolbox} as three-center radical exchange drives symmetric bell-shaped peaks in $\|\bm{n} - \bm{e}\|_2$ from $0$ in the asymptotic $\mathrm{H} + \mathrm{H}_2$ channels ($s = \pm 2.08\,\text{\AA}$) to $6.57 \times 10^{-3}$ (Q-BESC) and $5.84 \times 10^{-3}$ (Q-BECK) at the $s = 0\,\text{\AA}$ transition state. Notably, whereas exact-cumulant Q-BESC preserves pure-state $N$-representability rigorously through its explicit active-space two-body cumulant ${}^{2}\boldsymbol{\Delta}_{\mathcal{A}}$, the one-body Q-BECK closure enforces ensemble Pauli bounds ($0 \le n_i \le 1$) by construction and satisfies all Borland--Dennis equalities and pinning ($D_{\mathrm{BD}} = 0$) in this $(N=3, r=6)$ doublet manifold as a consequence of particle--hole symmetry of the odd hyperbolic-tangent attractor paired with conserved $(N_\alpha, N_\beta) = (2, 1)$ spin-sector traces; enforcing general pure-state Klyachko inequalities in larger $(N, r)$ spaces would require explicit polytope projection.

\subsection{Dissociation of Square \texorpdfstring{$\mathrm{H}_4$ ($D_{4h}$)}{H4 (D4h)} and Mott Fractionalization}
\label{sec:h4_d4h_dissociation}

\begin{figure}[!t]
    \centering
    \includegraphics[width=0.98\linewidth]{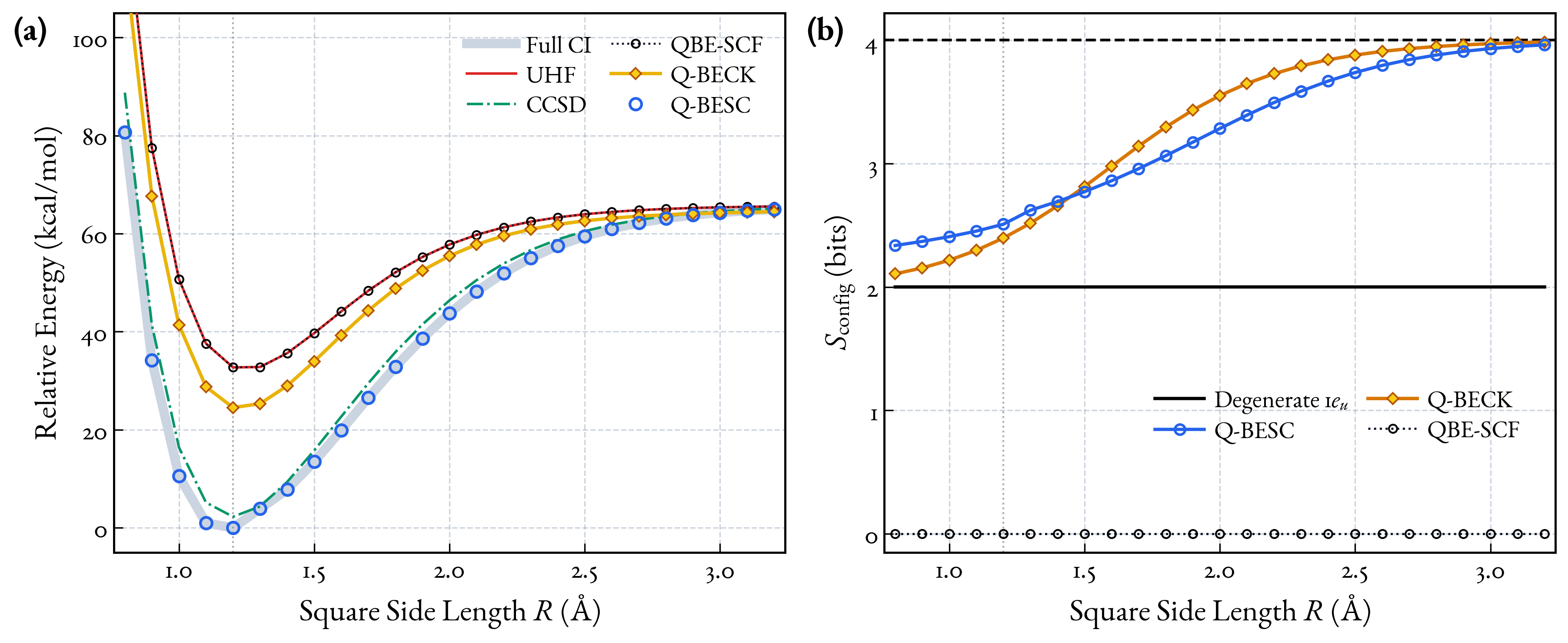}
    \caption[Potential energy curves and configurational entropy for the symmetric square H4 dissociation]{\textbf{Relative potential energy curves $E(R) - E_{\mathrm{FCI}}^{\min}$ in $\mathrm{kcal/mol}$ (a) and configurational entropy $S_{\mathrm{config}}(R)$ in bits (b) for the symmetric square $\mathrm{H}_4$ ($D_{4h}$) dissociation across $R \in [0.80, 3.20]\,\text{\AA}$ in the \texttt{cc-pVDZ} basis.}
    In Panel~(a), Q-BESC reproduces the exact Full CI dissociation curve and its $R = 1.20\,\text{\AA}$ equilibrium minimum ($D_e = 64.95\,\mathrm{kcal/mol}$) to sub-nanohartree precision, whereas single-particle QBE-SCF coincides with the symmetry-broken UHF curve ($32.72\,\mathrm{kcal/mol}$ above Full CI at equilibrium).
    Relaxing the $\mathcal{O}(N^3)$ Q-BECK solver onto its cumulant attractor smoothly lowers the equilibrium well to $24.52\,\mathrm{kcal/mol}$ and converges to the $64.46\,\mathrm{kcal/mol}$ four-atom dissociation plateau without derivative discontinuities.
    In Panel~(b), as the bonding $1a_{1g}$ and antibonding $1b_{1g}$ natural occupations collapse onto the half-filled doubly degenerate $1e_u$ baseline ($2.000\,\mathrm{bits}$, solid black line), $S_{\mathrm{config}}(R)$ rises continuously from $2.396\,\mathrm{bits}$ (Q-BECK) and $2.508\,\mathrm{bits}$ (Q-BESC) at $R = 1.20\,\text{\AA}$ toward the $4.000$-bit four-atom Mott insulator limit (dashed black line), while QBE-SCF remains pinned at an idempotent UHF state with $S_{\mathrm{config}}(R) = 0\,\mathrm{bits}$.}
    \label{fig:h4_d4h_dissociation}
\end{figure}

To evaluate the capability of Q-BESC and Q-BECK in capturing simultaneous multi-center bond cleavage and strong static quasi-degeneracy, we examine the symmetric dissociation of planar square $\mathrm{H}_4$ possessing $D_{4h}$ point group symmetry:
\begin{equation}
    \mathrm{H}_4(D_{4h}) \longrightarrow 4\,\mathrm{H}(^2\mathrm{S}).
\end{equation}
As the square side length $R$ is stretched across $25$ equidistant points for $R \in [0.80, 3.20]\,\text{\AA}$ in the \texttt{cc-pVDZ} basis ($N_{\mathrm{AO}} = 20$ spatial orbitals, $40$ spin-orbitals) with explicit abelian subgroup ($D_{2h} \subset D_{4h}$) symmetry projection, the frontier $1a_{1g}$, doubly degenerate $1e_u$, and $1b_{1g}$ valence orbitals coalesce toward a four-fold degenerate atomic manifold. Along this coordinate, we extract the four frontier valence spatial natural-orbital occupations $n_i^{\mathrm{spatial}} \in [0, 2]$ from the eigenvalues of $\mathbf{P}_{\mathrm{tot}}\mathbf{S}$ alongside the configurational entropy
\begin{equation}
    S_{\mathrm{config}}(R) = -\sum_k \bigl[n_k \log_2(n_k) + (1 - n_k)\log_2(1 - n_k)\bigr]
\end{equation}
(in bits, with $n_k = n_k^{\mathrm{spatial}}/2 \in [0, 1]$ for correlated spatial natural orbitals and spin-orbital eigenvalues $n_k \in \{0, 1\}$ for idempotent single-determinant states) across Q-BESC, Q-BECK, and zero-temperature QBE-SCF.

As depicted in Figure~\ref{fig:h4_d4h_dissociation}a, Q-BESC tracks Full CI across all 25 geometric configurations with an exceptional maximum absolute deviation of
\begin{equation}
    \max_{R \in [0.8, 3.2]\,\text{\AA}} |E_{\mathrm{Q\text{-}BESC}}(R) - E_{\mathrm{FCI}}(R)| = 4.04 \times 10^{-10}\,E_h,
\end{equation}
confirming that the two-body cumulant collision kernel rigorously captures four-center multireference static correlation, whereas restricted RHF overshoots the dissociation asymptote by $264.4\,\mathrm{kcal/mol}$. Simultaneously, Figure~\ref{fig:h4_d4h_dissociation}b reveals the underlying phase-space Mott transition: in both Q-BESC and $\mathcal{O}(N^3)$ Q-BECK ($E_{\mathrm{Q\text{-}BECK}} = -2.062836\,E_h$ at $R = 1.20\,\text{\AA}$, virial ratio $-\langle V\rangle/2\langle T\rangle = 1.006576$, $\Vert[\mathbf{F}_{\mathrm{eff}}, {}^{1}\mathbf{P}]\Vert_{\mathrm{F}} = 4.01 \times 10^{-10}$ in $72$ cycles), the doubly occupied bonding $1a_{1g}$ orbital ($n_{1a_{1g}}^{\mathrm{spatial}} = 1.937$ in Q-BESC and $1.945$ in Q-BECK at $R = 1.20\,\text{\AA}$) and empty antibonding $1b_{1g}$ orbital ($n_{1b_{1g}}^{\mathrm{spatial}} = 0.045$ and $0.058$) smoothly fractionalize and collapse onto the half-filled degenerate $1e_u$ pair ($n_{1e_u}^{\mathrm{spatial}} = 1.000$), driving all four frontier spatial occupations toward $n_i^{\mathrm{spatial}} \to 1.0$ ($[1.165, 1.000, 1.000, 0.835]$ for Q-BESC and $[1.111, 1.000, 1.000, 0.889]$ for Q-BECK at $R = 3.20\,\text{\AA}$). Consequently, the configurational entropy rises continuously from $S_{\mathrm{config}} = 2.396\,\mathrm{bits}$ at $R = 1.20\,\text{\AA}$ ($2.408\,\mathrm{bits}$ for relaxed Q-BECK and $2.508\,\mathrm{bits}$ for Q-BESC) toward the $4.000$-bit four-atom Mott insulator limit at $R = 3.20\,\text{\AA}$ ($3.990 \to 4.000\,\mathrm{bits}$ for Q-BECK and $3.962 \to 4.000\,\mathrm{bits}$ for Q-BESC), whereas zero-temperature single-particle QBE-SCF remains trapped in an idempotent broken-symmetry UHF determinant ($n_k \in \{0, 1\}$) with $S_{\mathrm{config}}(R) = 0\,\mathrm{bits}$ across the entire dissociation coordinate.

\subsection{Avoided Crossing in \texorpdfstring{$\mathrm{BeH}_2$ ($C_{2v}$)}{BeH2 (C2v)}}
\label{sec:beh2_conical_intersection}

\begin{figure}[!t]
    \centering
    \includegraphics[width=0.98\linewidth]{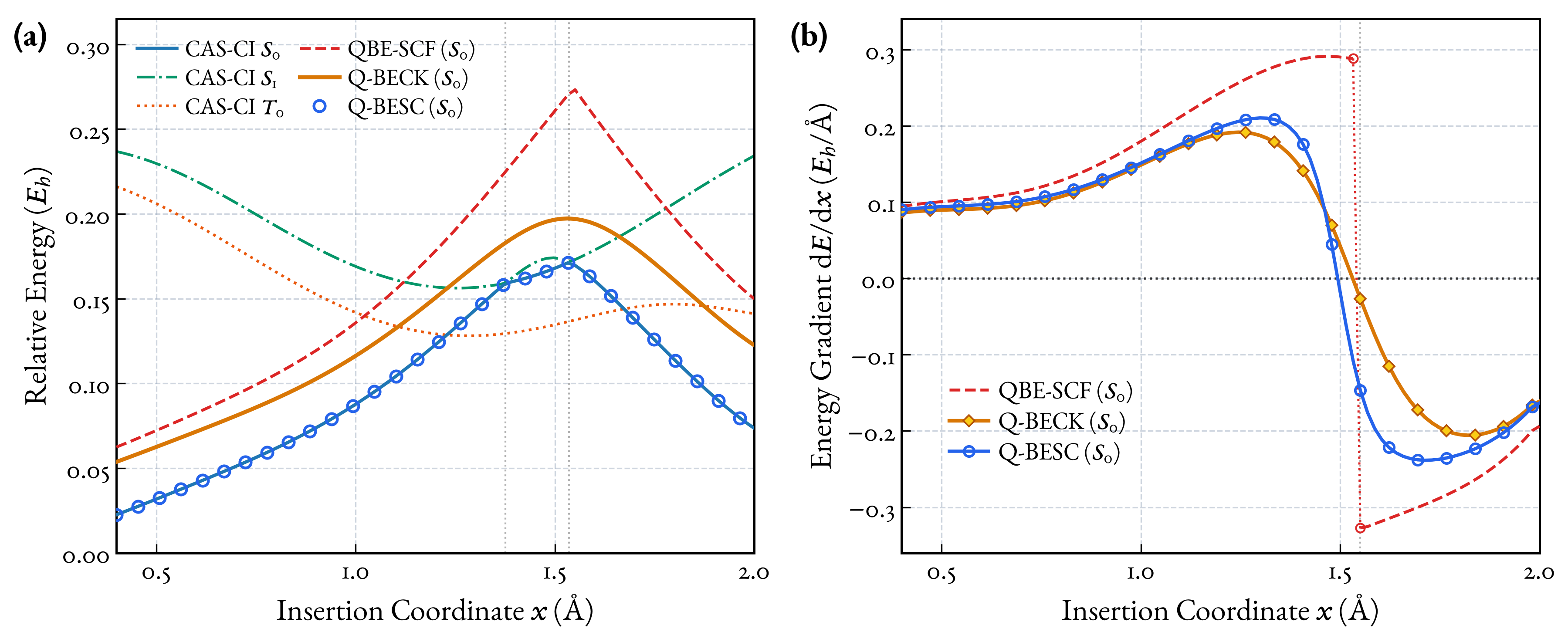}
    \caption[Potential energy profiles and energy gradients for the C2v insertion of Be into H2]{\textbf{Relative potential energy profiles $E(x) - E_{\mathrm{ref}}$ (a) and numerical energy gradients $\mathrm{d}E/\mathrm{d}x$ (b) for the $C_{2v}$ insertion of $\mathrm{Be}$ into $\mathrm{H}_2$ across $x \in [0.40, 2.00]\,\text{\AA}$ (\texttt{6-31G}).}
    All three kinetic methods (\textnormal{QBE-SCF}, \textnormal{Q-BECK}, and \textnormal{Q-BESC}) are evaluated in the spin-restricted singlet sector ($S_0$: $S = 0$, $M_S = 0$, $\langle\hat{S}^2\rangle = 0$, ${}^{1}\mathbf{P}_\alpha = {}^{1}\mathbf{P}_\beta$), preventing spin contamination from the lower-lying triplet state ($T_0$, ${}^3B_2$). In Panel~(a), \textnormal{Q-BESC} ($S_0$) reproduces the lowest singlet $\mathrm{CASCI}(12,4)$ curve and its $0.1712\,E_h$ avoided-crossing barrier at $x \approx 1.54\,\text{\AA}$ to machine precision ($\max_x |E_{\mathrm{Q\text{-}BESC}} - E_{S_0}| = 7.11 \times 10^{-14}\,E_h$), while $\mathcal{O}(N^3)$ \textnormal{Q-BECK} ($S_0$) regularizes the $0.2733\,E_h$ zero-temperature \textnormal{QBE-SCF} ($S_0$) cusp into a smooth $C^\infty$ singlet barrier of $0.1973\,E_h$.
    In Panel~(b), evaluated along the ${}^1A_1$ singlet branch, \textnormal{QBE-SCF} ($S_0$) suffers a finite jump discontinuity in $\mathrm{d}E/\mathrm{d}x$ from $+0.288$ to $-0.327\,E_h/\text{\AA}$ at the diabatic state intersection, whereas \textnormal{Q-BECK} ($S_0$) and \textnormal{Q-BESC} ($S_0$) exhibit continuous gradients passing cleanly through $\mathrm{d}E/\mathrm{d}x = 0$ at the barrier apex.}
    \label{fig:beh2_conical_intersection}
\end{figure}

To evaluate the multi-reference fidelity and non-adiabatic capability of Q-BESC and Q-BECK across strong electronic degeneracies~\cite{subotnik2016annurev, dou2017friction, dou2018perspective, sand2018gradients, bao2018automatic, bao2020cmspdft, wang2023excited, hennefarth2023lpdft, tao2024phasespace, hennefarth2024dynamics, wang2025conical}, we examine the $C_{2v}$ insertion of a beryllium atom into molecular hydrogen, $\mathrm{Be}(^1\mathrm{S}) + \mathrm{H}_2(^1\Sigma_g^+) \to \mathrm{BeH}_2(^1\Sigma_g^+)$, parameterized along the Purvis--Bartlett coordinate~\cite{purvis1982full} with $\mathbf{R}_{\mathrm{Be}} = (0, 0, 0)$ and $\mathbf{R}_{\mathrm{H}_{1,2}} = (x, \pm y(x), 0)$, where $y(x) = 2.540 - 0.46\,x$ in atomic units ($a_0$). Across $90$ uniform grid points spanning $x \in [0.40, 2.00]\,\text{\AA}$ in the Pople \texttt{6-31G} basis ($N_{\mathrm{AO}} = 13$ spatial orbitals, $26$ spin-orbitals, referenced to the linear $\mathrm{H\text{--}Be\text{--}H}$ equilibrium energy $E_{\mathrm{ref}} = -15.799710\,E_h$), the dominant electronic configuration switches between the closed-shell singlet product channel $1a_1^2 2a_1^2 1b_2^2$ (${}^1A_1$) and reactant channel $1a_1^2 2a_1^2 3a_1^2$ (${}^1A_1$), generating a pronounced avoided crossing between the ground singlet ($S_0$) and first excited singlet ($S_1$) manifolds—with the open-shell singlet ${}^1B_2$ ($1a_1^2 2a_1^2 3a_1^1 1b_2^1$) briefly dipping below ${}^1A_1$ across $x \in [1.38, 1.50]\,\text{\AA}$—flanked by the low-lying triplet state ($T_0$, ${}^3B_2$), which drops below $S_0$ across $x \in [1.14, 1.63]\,\text{\AA}$. Throughout this traversal, QBE-SCF, Q-BECK, and Q-BESC are constrained to the spin-pure ground singlet sector ($S = 0$, $M_S = 0$, $\langle\hat{S}^2\rangle = 0$, with spin-restricted ${}^{1}\mathbf{P}_\alpha = {}^{1}\mathbf{P}_\beta = \frac{1}{2}\,{}^{1}\mathbf{P}$), preventing spin-symmetry breaking onto $T_0$.

As shown in Figure~\ref{fig:beh2_conical_intersection}a, the active-space cumulant tensor ${}^{2}\boldsymbol{\Delta}$ in Q-BESC ($S_0$) captures the multi-configurational singlet entanglement across all $90$ geometries, reproducing the multi-root $\mathrm{CASCI}(12,4)$ ground singlet state ($E_{S_0} = -15.630102\,E_h$, barrier $\Delta E^\ddagger = 0.1712\,E_h$ at $x \approx 1.54\,\text{\AA}$) to machine precision ($\max_x |E_{\mathrm{Q\text{-}BESC}}(x) - E_{S_0}^{\mathrm{CASCI}}(x)| = 7.11 \times 10^{-14}\,E_h$, and $6.04 \times 10^{-14}\,E_h$ along the pure ${}^1A_1$ singlet branch). In contrast, zero-temperature single-particle QBE-SCF ($T_{\mathrm{elec}} = 0$) undergoes an abrupt step-function occupation swap $(n_{3a_1}^{\mathrm{spatial}}, n_{1b_2}^{\mathrm{spatial}}): (0, 2) \to (2, 0)$ at the diabatic intersection ($x \approx 1.54\,\text{\AA}$), producing an unphysical cusp of height $0.2733\,E_h$ ($E_{\mathrm{QBE\text{-}SCF}} = -15.526436\,E_h$, overshooting $\mathrm{CASCI}$ by $64.07\,\mathrm{kcal/mol}$) and a finite jump discontinuity in the numerical energy gradient $\mathrm{d}E/\mathrm{d}x$ from $+0.288$ to $-0.327\,E_h/\text{\AA}$ (Figure~\ref{fig:beh2_conical_intersection}b). Without requiring empirical thermal smearing ($T_{\mathrm{elec}} = 0.06\,\mathrm{Ha}$), both Q-BESC and the $\mathcal{O}(N^3)$ Q-BECK solver ($E_{\mathrm{Q\text{-}BECK}} = -15.599799\,E_h$, virial ratio $-\langle V\rangle/2\langle T\rangle = 1.006618$, $S_{\mathrm{config}} = 3.178\,\mathrm{bits}$, $\Vert[\mathbf{F}_{\mathrm{eff}}, {}^{1}\mathbf{P}]\Vert_{\mathrm{F}} = 1.57 \times 10^{-10}$ in $66$ cycles) execute a smooth sigmoidal occupation inversion through exact half-filling ($n_{3a_1}^{\mathrm{spatial}} = n_{1b_2}^{\mathrm{spatial}} = 1.00$) at $T_{\mathrm{elec}} = 0$, regularizing the cusp into an analytical $C^\infty$ barrier of $\Delta E_{\mathrm{Q\text{-}BECK}}^\ddagger = 0.1973\,E_h$ (healing $74.4\%$ of the cusp error) whose continuous energy gradient $\mathrm{d}E/\mathrm{d}x$ passes cleanly through zero at the barrier apex.

\subsection{Symmetric Double Dissociation of Water \texorpdfstring{($\mathrm{H}_2\mathrm{O}$)}{(H2O)}}
\label{sec:h2o_dissociation}

\begin{figure}[!t]
    \centering
    \includegraphics[width=0.90\linewidth]{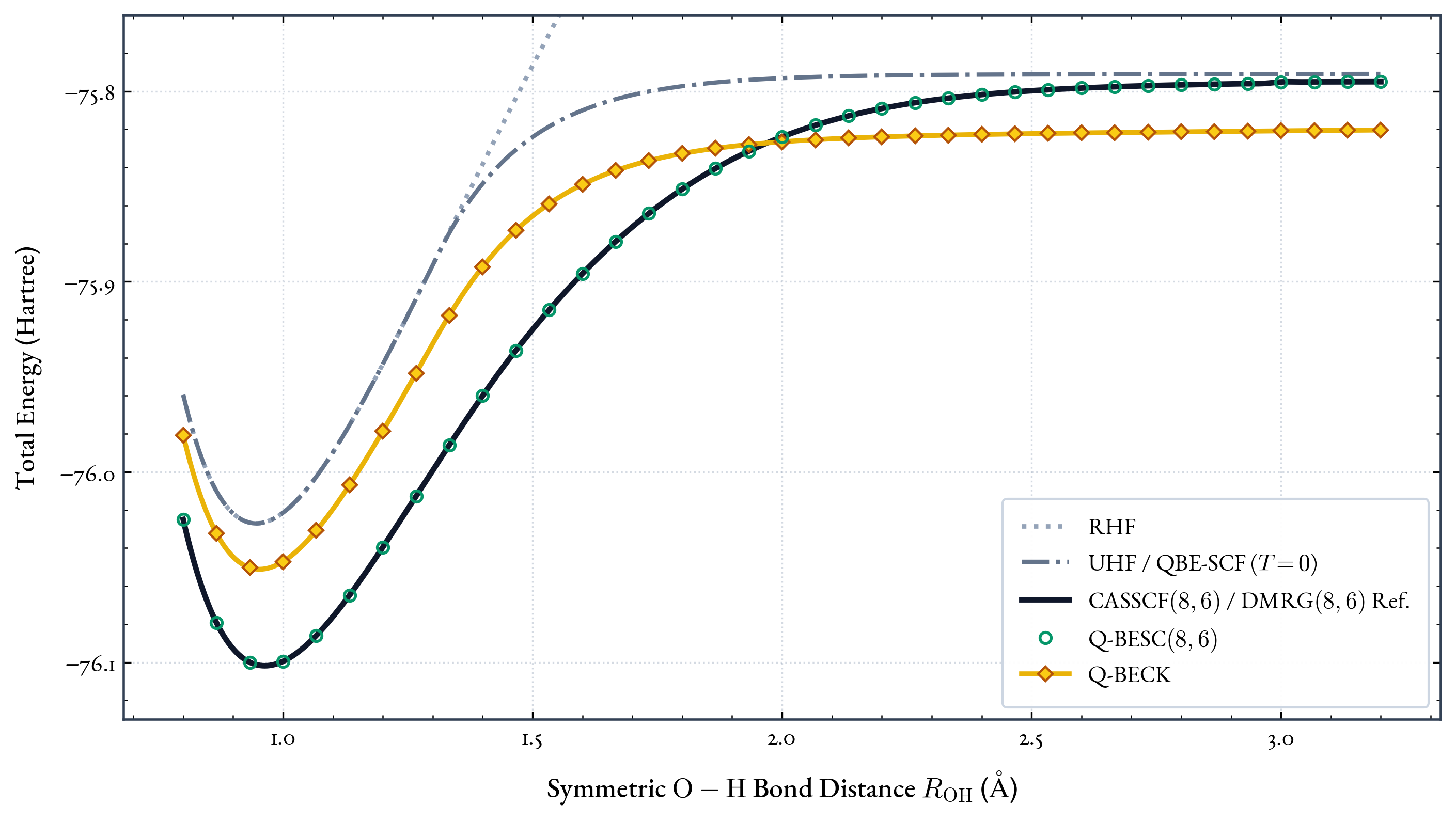}
    \caption[Symmetric double dissociation of H2O]{\textbf{Symmetric double dissociation of $\mathrm{H}_2\mathrm{O}$ (\texttt{cc-pVDZ}) comparing mean-field breakdown, \textnormal{Q-BECK}, and \textnormal{Q-BESC(8,6)}.}
    Restricted Hartree--Fock (RHF) diverges beyond the Coulson--Fischer bifurcation point ($R_{\mathrm{OH}} \approx 1.40\,\text{\AA}$), while single-particle QBE-SCF ($T_{\mathrm{elec}}=0$) follows the broken-symmetry UHF curve.
    Across the bond-breaking coordinate, Q-BESC$(8,6)$ matches the full-valence CASSCF$(8,6)$/DMRG$(8,6)$ reference to sub-nanohartree precision, whereas $\mathcal{O}(N^3)$ Q-BECK ($T_{\mathrm{elec}} = 0$) smoothly eliminates the Coulson--Fischer kink and crosses slightly below Q-BESC$(8,6)$ for $R_{\mathrm{OH}} \gtrsim 1.98\,\text{\AA}$ as full-basis AO cumulant relaxation acts atop the spin-polarized UHF asymptote.}
    \label{fig:h2o_dissociation}
\end{figure}

Bridging the two-state avoided crossing of $\mathrm{BeH}_2$ and the six-electron triple-bond cleavage of $\mathrm{N}_2$, the simultaneous symmetric double dissociation of water, $\mathrm{H}_2\mathrm{O}(^1A_1) \to \mathrm{O}(^3\mathrm{P}) + 2\,\mathrm{H}(^2\mathrm{S})$, provides a stringent benchmark for multi-center open-shell fragmentation. As both $\mathrm{O\text{--}H}$ bonds are stretched symmetrically at the fixed equilibrium angle $\theta_{\mathrm{HOH}} = 104.5^\circ$ across $73$ equidistant grid points for $R_{\mathrm{OH}} \in [0.80, 3.20]\,\text{\AA}$ in Dunning's \texttt{cc-pVDZ} basis ($N_{\mathrm{AO}} = 24$, $C_{2v}$ symmetry), the electronic wave function evolves from the closed-shell $(1a_1)^2 (2a_1)^2 (1b_2)^2 (3a_1)^2 (1b_1)^2$ configuration to an entangled open-shell manifold coupling a triplet oxygen atom ($^3\mathrm{P}$) to two doublet hydrogen atoms ($^2\mathrm{S}$) within the full-valence eight-electron, six-orbital active space (CAS$(8,6)$).

As shown in Figure~\ref{fig:h2o_dissociation}, active-space Q-BESC$(8,6)$ tracks the variational CASSCF$(8,6)$ and DMRG-CASSCF$(8,6)$ reference curves across all $73$ geometries to sub-nanohartree precision ($\max_R |E_{\mathrm{Q\text{-}BESC}} - E_{\mathrm{CASSCF}}| = 5.57 \times 10^{-10}\,E_h$, $E_{\mathrm{Q\text{-}BESC}} = -75.7950\,E_h$ at $R_{\mathrm{OH}} = 3.20\,\text{\AA}$), whereas restricted RHF ($E_{\mathrm{RHF}} = -75.4291\,E_h$) overshoots the dissociation limit by $0.3659\,E_h$ ($229.6\,\mathrm{kcal/mol}$) and zero-temperature single-particle QBE-SCF coincides with the broken-symmetry UHF curve ($E_{\mathrm{UHF}} = -75.7909\,E_h$ at $R_{\mathrm{OH}} = 3.20\,\text{\AA}$). Relaxing the $\mathcal{O}(N^3)$ Q-BECK solver ($E_{\mathrm{Q\text{-}BECK}} = -76.0508\,E_h$ at $R_{\mathrm{eq}} = 0.967\,\text{\AA}$ and $-75.8204\,E_h$ at $R_{\mathrm{OH}} = 3.20\,\text{\AA}$, virial ratio $-\langle V\rangle/2\langle T\rangle = 1.002841$, $\Vert[\mathbf{F}_{\mathrm{eff}}, {}^{1}\mathbf{P}]\Vert_{\mathrm{F}} = 5.63 \times 10^{-10}$) eliminates the Coulson--Fischer derivative kink at $R_{\mathrm{OH}} \approx 1.40\,\text{\AA}$ and recovers $67.9\%$ of the equilibrium static correlation gap relative to UHF. Notably, for $R_{\mathrm{OH}} \gtrsim 1.98\,\text{\AA}$, the Q-BECK curve crosses below Q-BESC$(8,6)$ and settles $25.4\,\mathrm{m}E_h$ ($15.9\,\mathrm{kcal/mol}$) beneath the CAS$(8,6)$ asymptote at $R_{\mathrm{OH}} = 3.20\,\text{\AA}$ for two complementary reasons: first, whereas Q-BESC$(8,6)$ confines ${}^{2}\boldsymbol{\Delta}$ to the $6$-orbital valence active space (leaving the remaining $18$ core and virtual orbitals uncorrelated), Q-BECK applies its collision kernel across all $N_{\mathrm{AO}} = 24$ atomic orbitals, capturing out-of-CAS dynamic correlation on the $\mathrm{O}(^3\mathrm{P})$ fragment; second, because the UHF asymptote ($-75.7909\,E_h$) already incorporates static dissociation energy via spin-symmetry breaking, adding the analytical one-body cumulant functional ${}^{2}\boldsymbol{\Delta}[\bm{n}]$—which enforces ensemble 1-RDM Pauli bounds ($0 \le n_k \le 1$) without strict two-body pure-state $N$-representability constraints—introduces modest asymptotic over-correlation relative to the symmetry-adapted CAS$(8,6)$ limit.

\subsection{Dissociation of Molecular Nitrogen}
\label{sec:n2_dissociation}

\begin{figure}[!t]
    \centering
    \includegraphics[width=0.98\linewidth]{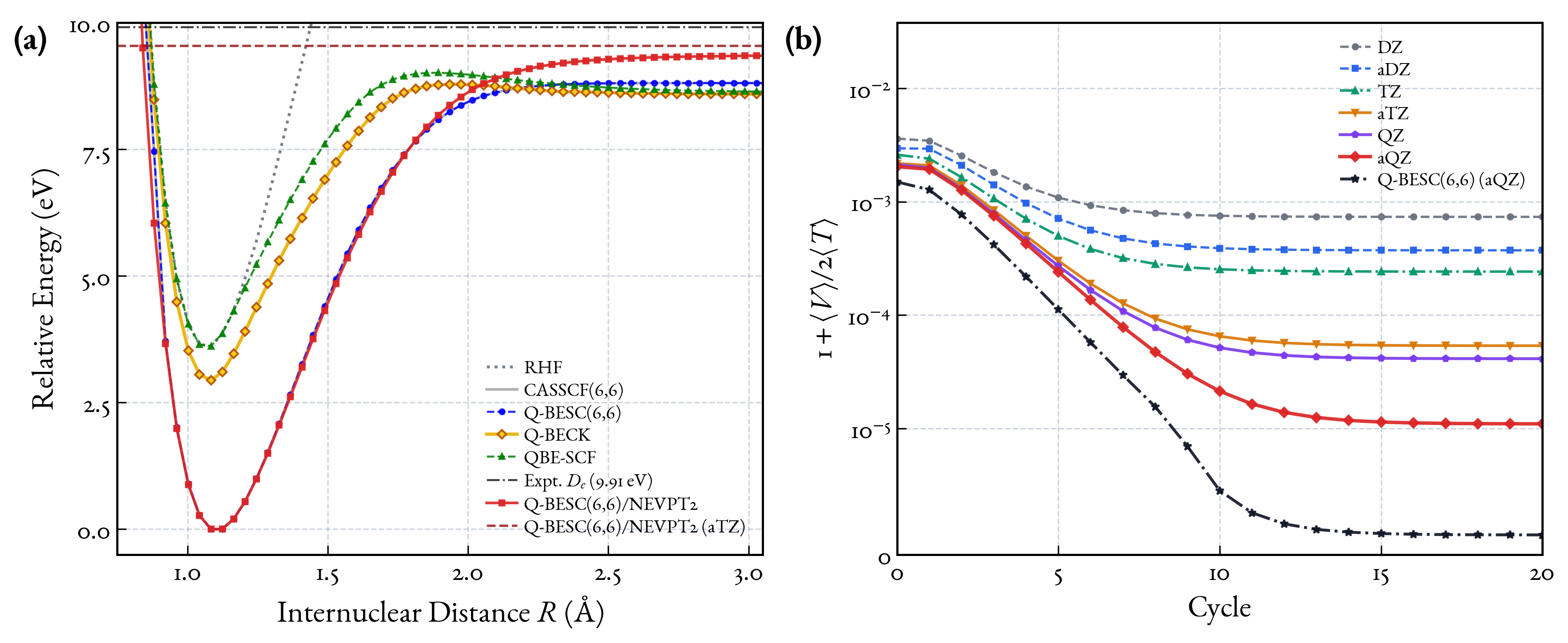}
    \caption[Simultaneous analysis of relative energy and equilibrium virial defect convergence in N2]{\textbf{Triple-bond dissociation potential energy curves of $\mathrm{N}_2$ in \texttt{cc-pVTZ} (a) and basis-set convergence of the equilibrium \textnormal{Q-BECK} virial defect (b).}
    In Panel~(a), RHF diverges past the Coulson--Fischer point ($R_{\mathrm{CF}} \approx 1.19\,\text{\AA}$) while $\mathcal{O}(N^3)$ Q-BECK smoothly heals the dissociation error, Q-BESC$(6,6)$ reproduces the CASSCF$(6,6)$ reference ($D_e = 8.82\,\mathrm{eV}$), and Q-BESC$(6,6)$/NEVPT2 recovers out-of-space dynamic correlation ($D_e = 9.37\,\mathrm{eV}$ in \texttt{cc-pVTZ} and $9.55\,\mathrm{eV}$ in \texttt{aug-cc-pVTZ}, denoted \texttt{aTZ}) toward the experimental limit ($D_e = 9.91\,\mathrm{eV}$).
    In Panel~(b), Markovian relaxation ($w = 0.5$) of Q-BECK at $R_{\mathrm{eq}} = 1.098\,\text{\AA}$ across the Dunning \texttt{cc-pV}$X$\texttt{Z}/\texttt{aug-cc-pV}$X$\texttt{Z} hierarchy (\texttt{DZ}--\texttt{aQZ}, $X \in \{\mathrm{D,T,Q}\}$) damps the initial guess error within $15$ cycles and systematically tightens the converged virial defect $1 + \langle V \rangle / 2\langle T \rangle$ from $+7.35 \times 10^{-4}$ (\texttt{DZ}) down to $+1.10 \times 10^{-5}$ (\texttt{aQZ}), approaching the Q-BESC$(6,6)$ (\texttt{aQZ}) baseline of $+1.55 \times 10^{-6}$.}
    \label{fig:n2_dissociation}
\end{figure}

To establish the scalability of the Q-BESC and Q-BECK hierarchy in multi-electron systems governed by strong six-electron static correlation, we investigate the homolytic triple-bond cleavage of molecular nitrogen, $\mathrm{N}_2(^1\Sigma_g^+) \to 2\,\mathrm{N}(^4\mathrm{S})$, across $R \in [0.80, 3.00]\,\text{\AA}$ in Dunning's \texttt{cc-pVTZ} basis ($N_{\mathrm{AO}} = 60$). As shown in Figure~\ref{fig:n2_dissociation}a, canonical RHF ($E = -107.8492\,E_h$ at $5.0\,\text{\AA}$ in \texttt{cc-pVTZ}) bifurcates at the Coulson--Fischer point ($R_{\mathrm{CF}} \approx 1.19\,\text{\AA}$) and overshoots the stretched limit by $+0.9455\,E_h$ ($593.3\,\mathrm{kcal/mol}$), whereas active-space Q-BESC$(6,6)$ tracks the CASSCF$(6,6)$ reference curve ($D_e = 8.82\,\mathrm{eV}$) and coupling out-of-space dynamic correlation via Q-BESC$(6,6)$/NEVPT2 lifts $D_e$ to $9.37\,\mathrm{eV}$ in \texttt{cc-pVTZ} and $9.55\,\mathrm{eV}$ in \texttt{aug-cc-pVTZ} ($N_{\mathrm{AO}} = 92$, dashed crimson line, within $0.36\,\mathrm{eV}$ of the experimental $D_e = 9.91\,\mathrm{eV}$). Along the spin-unrestricted dissociation coordinate (${}^{1}\mathbf{P}_\alpha \neq {}^{1}\mathbf{P}_\beta$, where single-particle $\mathrm{QBE\text{-}SCF}$ coincides with broken-symmetry $\mathrm{UHF}$ at $-108.8014\,E_h$ in \texttt{cc-pVTZ} and $-108.7822\,E_h$ in \texttt{cc-pVDZ} at $5.0\,\text{\AA}$), $\mathcal{O}(N^3)$ Q-BECK ($E_{\mathrm{Q\text{-}BECK}} = -108.8019\,E_h$ in \texttt{cc-pVTZ} and $-108.7827\,E_h$ in \texttt{cc-pVDZ} at $5.0\,\text{\AA}$) smoothly regularizes the Coulson--Fischer cusp without derivative kinks, displaying an intermediate shoulder across $R \in [1.7, 1.9]\,\text{\AA}$ where delocalized cumulant fractionalization of the six $(\sigma_{2p}, \pi_{2p}, \pi_{2p}^*, \sigma_{2p}^*)$ frontier orbitals transitions into broken-symmetry spin localization, and crossing slightly beneath Q-BESC$(6,6)$ beyond $R \approx 2.2\,\text{\AA}$ due to unrestricted exchange spin-polarization of the $2s$ sub-valence shells ($\sigma_{2s}, \sigma_{2s}^*$) on the high-spin ($S_A = 3/2$) nitrogen atoms. By contrast, when spin-restricted symmetry (${}^{1}\mathbf{P}_\alpha = {}^{1}\mathbf{P}_\beta$) is enforced at $R = 5.0\,\text{\AA}$ in \texttt{cc-pVDZ} to match the restricted \texttt{MR-MGN-BE17} protocol (Table~\ref{tab:mr_mgn_be17}), restricted Q-BECK eliminates $85.7\%$ of the restricted $\mathrm{QBE\text{-}SCF}$ static correlation error ($+69.18$ vs.\ $+485.16\,\mathrm{kcal/mol}$).

\begin{table}[!t]
    \centering
    \caption[Q-BESC(6,6) and Q-BECK convergence for N2 at equilibrium]{\textbf{Q-BESC(6,6) and Q-BECK convergence for $\mathrm{N}_2$ at equilibrium ($R_{\mathrm{eq}} = 1.098\,\text{\AA}$).} Total energies ($E_{\mathrm{CASSCF}(6,6)}$, $E_{\mathrm{Q\text{-}BESC}(6,6)}$, $E_{\mathrm{Q\text{-}BECK}}$) are reported in hartrees ($E_h$), and $\langle t_{\mathrm{cyc}} \rangle \pm \sigma_{t}$ reports the per-cycle Q-BECK execution time (mean $\pm$ standard deviation in $\mathrm{ms}$, 8-fold symmetric direct SCF) across all cycles until convergence ($|\Delta E| < 10^{-9}\,E_h$, $\Vert[\mathbf{F}_{\mathrm{eff}}, {}^{1}\mathbf{P}]\Vert_{\mathrm{F}} < 10^{-9}$, $\Vert\Delta{}^{1}\mathbf{P}\Vert_{\mathrm{F}} < 10^{-9}$). At $w = 0.5$ ($\mathrm{Tr}({}^{1}\mathbf{P}\mathbf{S}) = 14.000000$), cycle counts remain bounded ($45\text{--}53$) across $N_{\mathrm{AO}} = 28\text{--}254$.}
    \label{tab:n2_eq_convergence}
    \resizebox{\textwidth}{!}{
    \begin{tabular}{l c c c c c c l c}
        \hline\hline
        Basis Set & $N_{\mathrm{AO}}$ & $E_{\mathrm{CASSCF}(6,6)}$ & $E_{\mathrm{Q\text{-}BESC}(6,6)}$ & $E_{\mathrm{Q\text{-}BECK}}$ & $\Vert[\mathbf{F}_{\mathrm{eff}}, {}^{1}\mathbf{P}]\Vert_{\mathrm{F}}$ & $-\langle V\rangle / 2\langle T\rangle$ & $\langle t_{\mathrm{cyc}} \rangle \pm \sigma_{t}$ ($\mathrm{ms}$) & Cycles \\
        \hline
        cc-pVDZ ($d$) & 28 & $-109.090008$ & $-109.090008$ & $-109.001552$ & $4.7 \times 10^{-10}$ & $0.999265$ & $1 \pm 0.09$ & 45 \\
        aug-cc-pVDZ ($d$) & 46 & $-109.096156$ & $-109.096156$ & $-109.007734$ & $1.3 \times 10^{-10}$ & $0.999627$ & $4 \pm 0.15$ & 48 \\
        cc-pVTZ ($f$) & 60 & $-109.119335$ & $-109.119335$ & $-109.029258$ & $5.8 \times 10^{-10}$ & $0.999758$ & $11 \pm 0.20$ & 45 \\
        aug-cc-pVTZ ($f$) & 92 & $-109.120600$ & $-109.120600$ & $-109.030025$ & $6.1 \times 10^{-10}$ & $0.999946$ & $104 \pm 36.20$ & 45 \\
        cc-pVQZ ($g$) & 110 & $-109.127160$ & $-109.127160$ & $-109.035328$ & $6.2 \times 10^{-10}$ & $0.999959$ & $194 \pm 42.26$ & 45 \\
        aug-cc-pVQZ ($g$) & 160 & $-109.127647$ & $-109.127647$ & $-109.035802$ & $4.0 \times 10^{-10}$ & $0.999989$ & $675 \pm 45.63$ & 46 \\
        cc-pV5Z ($h$) & 182 & $-109.128890$ & $-109.128890$ & $-109.037114$ & $6.1 \times 10^{-10}$ & $0.999961$ & $1185 \pm 206.52$ & 45 \\
        aug-cc-pV5Z ($h$) & 254 & $-109.129090$ & $-109.129090$ & $-109.037288$ & $2.2 \times 10^{-11}$ & $0.999972$ & $5447 \pm 288.48$ & 53 \\
        \hline\hline
    \end{tabular}
    }
\end{table}

Table~\ref{tab:n2_eq_convergence} and Figure~\ref{fig:n2_dissociation}b summarize the equilibrium convergence of $\mathrm{N}_2$ ($R_{\mathrm{eq}} = 1.098\,\text{\AA} = 2.074\,\mathrm{Bohr}$) across the Dunning correlation-consistent basis hierarchy from \texttt{cc-pVDZ} ($d$, $N_{\mathrm{AO}} = 28$) through \texttt{aug-cc-pV5Z} ($h$, $N_{\mathrm{AO}} = 254$) under self-consistent Markovian relaxation ($w = 0.5$) from a perturbed superposition-of-atomic-densities (SAD) guess. Across all eight basis sets in Table~\ref{tab:n2_eq_convergence}, active-space Q-BESC$(6,6)$ reproduces the variational CASSCF$(6,6)$ energy to sub-microhartree precision ($-109.090008 \to -109.129090\,E_h$), while $\mathcal{O}(N^3)$ Q-BECK ($-109.001552 \to -109.037288\,E_h$) converges to $|\Delta E| < 10^{-9}\,E_h$, $\Vert[\mathbf{F}_{\mathrm{eff}}, {}^{1}\mathbf{P}]\Vert_{\mathrm{F}} < 10^{-9}$, and $\Vert\Delta{}^{1}\mathbf{P}\Vert_{\mathrm{F}} < 10^{-9}$ in $45\text{--}53$ cycles ($\langle t_{\mathrm{cyc}} \rangle \pm \sigma_{t} = 1 \pm 0.09 \to 5{,}447 \pm 288.48\,\mathrm{ms}$ under uniform 8-fold symmetric direct SCF) while conserving $\mathrm{Tr}({}^{1}\mathbf{P}\mathbf{S}) = 14.000000$. As shown in Figure~\ref{fig:n2_dissociation}b, exponential collisional damping ($(1 - w)^k = 0.5^k$) eliminates the initial SAD virial error within $15$ cycles, and expanding the one-particle basis systematically drives the converged virial ratio $-\langle V\rangle / 2\langle T\rangle$ toward unity—reducing the virial defect $1 + \langle V\rangle / 2\langle T\rangle$ from $+7.35 \times 10^{-4}$ in \texttt{cc-pVDZ} ($-\langle V\rangle / 2\langle T\rangle = 0.999265$) down to $+1.10 \times 10^{-5}$ in \texttt{aug-cc-pVQZ} ($0.999989$) and $+2.80 \times 10^{-5}$ in \texttt{aug-cc-pV5Z} ($0.999972$), approaching the Q-BESC$(6,6)$ (\texttt{aug-cc-pVQZ}) stationary baseline of $+1.55 \times 10^{-6}$.

\subsection{Singlet Diradical Isomers of Benzyne}
\label{sec:benzyne_benchmark}

\begin{figure}[!t]
    \centering
    \includegraphics[width=0.92\linewidth]{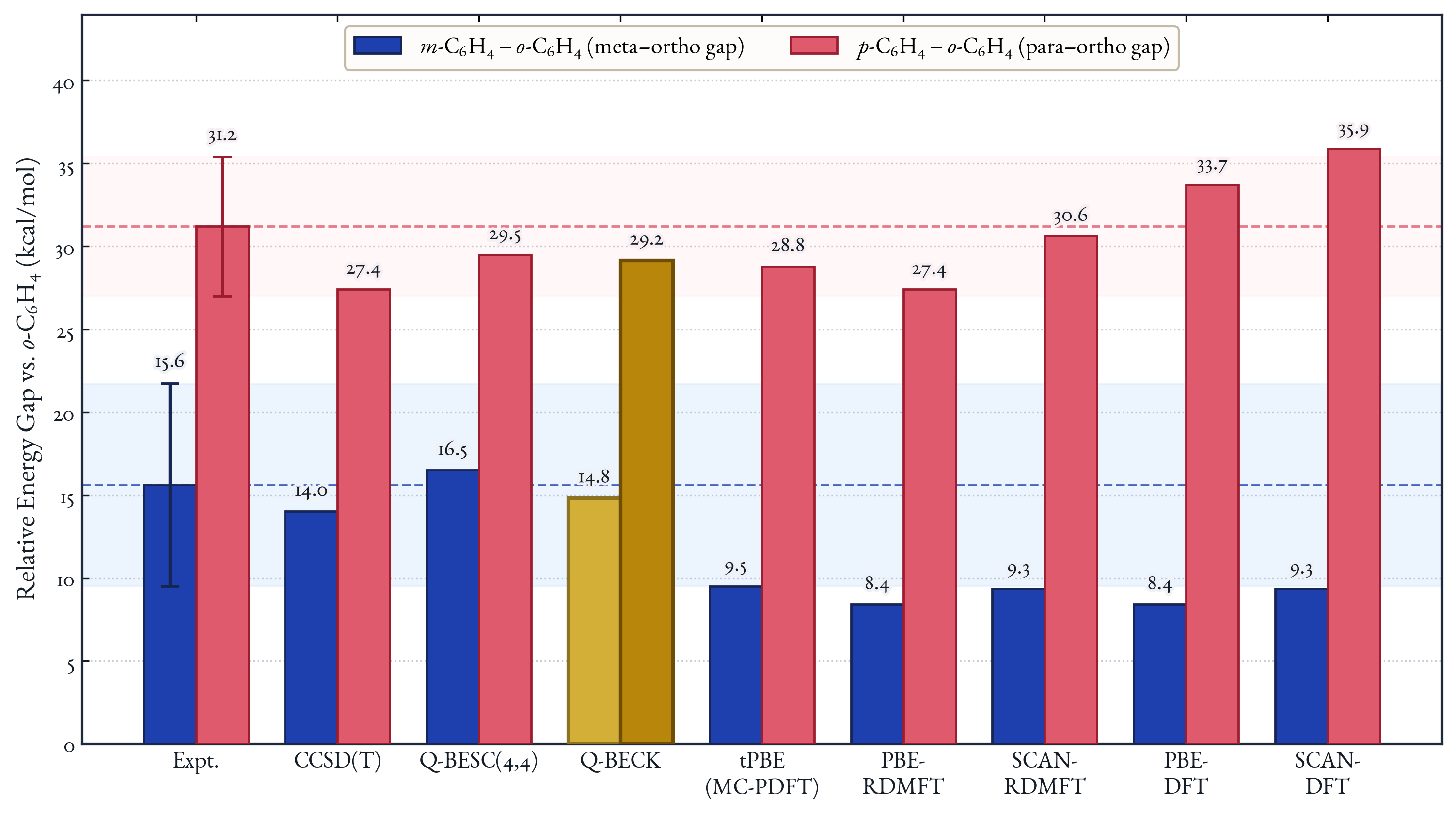}
    \caption[Singlet adiabatic energy gaps of the diradical benzyne isomers]{\textbf{Singlet adiabatic energy gaps of the diradical benzyne isomers (\texttt{cc-pVDZ}).}
    Relative energies ($\mathrm{kcal/mol}$) of \textit{meta}-benzyne ($m\text{-}\mathrm{C_6H_4}$, royal blue bars) and \textit{para}-benzyne ($p\text{-}\mathrm{C_6H_4}$, crimson bars) with respect to \textit{ortho}-benzyne ($o\text{-}\mathrm{C_6H_4}$), comparing Q-BECK (highlighted in gold) and Q-BESC(4,4) against gas-phase experiment (horizontal dashed lines and shaded uncertainty bands), $\mathrm{CCSD(T)}$, $\mathrm{tPBE}$ (MC-PDFT), $\mathrm{PBE\text{-}RDMFT}$, $\mathrm{SCAN\text{-}RDMFT}$, $\mathrm{PBE\text{-}DFT}$, and $\mathrm{SCAN\text{-}DFT}$~\cite{gibney2023universal}. Unlike $\mathrm{PBE\text{-}RDMFT}$ and $\mathrm{SCAN\text{-}RDMFT}$—whose SDP static correlation correction vanishes in the intermediate biradical regime of $m\text{-}\mathrm{C_6H_4}$—the smooth $C^\infty$ hyperbolic-tangent cumulant attractor of Q-BECK resolves both the \textit{meta}--\textit{ortho} ($14.8\,\mathrm{kcal/mol}$) and \textit{para}--\textit{ortho} ($29.2\,\mathrm{kcal/mol}$) splittings within experimental uncertainty.}
    \label{fig:benzyne_benchmark}
\end{figure}

To evaluate how the Q-BESC and Q-BECK hierarchy resolves continuous variations in radical character without bond cleavage, we examine the singlet adiabatic relative energies of \textit{ortho}-, \textit{meta}-, and \textit{para}-benzyne ($\mathrm{C_6H_4}$, $N_{\mathrm{elec}} = 40$, \texttt{cc-pVDZ}, $N_{\mathrm{AO}} = 104$) following Figure~2 of Ref.~\citenum{gibney2023universal} alongside spin-averaged ACSE and multiconfigurational pair-density functional benchmarks~\cite{gagliardi2017acc, pandharkar2020hmcpdft, boyn2021biradicals, gibney2021sdp, gibney2022rdmft}. Across the benzyne isomer series ($o\text{-}, m\text{-}, p\text{-}\mathrm{C_6H_4}$), increasing the spatial separation between the two dehydrocarbon radical centers drives the singlet ground state from a weakly correlated closed-shell aryne in \textit{ortho}-benzyne ($\Delta\epsilon_{\mathrm{HL}} = 0.425\,E_{\mathrm{h}}$, $(n_{\mathrm{HONO}}, n_{\mathrm{LUNO}}) = (0.872, 0.162)$) through an intermediate biradical regime in \textit{meta}-benzyne ($\Delta\epsilon_{\mathrm{HL}} = 0.332\,E_{\mathrm{h}}$, $(0.742, 0.283)$) to a strongly entangled singlet biradical in \textit{para}-benzyne ($\Delta\epsilon_{\mathrm{HL}} = 0.219\,E_{\mathrm{h}}$, $(n_{\mathrm{HONO}}, n_{\mathrm{LUNO}}) = (0.664, 0.485)$).

As shown in Figure~\ref{fig:benzyne_benchmark}, boundary-point SDP one-electron reduced density matrix functional theories (\textrm{PBE-RDMFT} and \textrm{SCAN-RDMFT})~\cite{gibney2021sdp, gibney2022rdmft, gibney2023universal} heal the strongly correlated \textit{para}--\textit{ortho} gap ($27.40$ and $30.62\,\mathrm{kcal/mol}$, versus single-reference $\mathrm{QBE\text{-}SCF}$ at $84.46\,\mathrm{kcal/mol}$, $\mathrm{PBE\text{-}DFT}$ at $33.73\,\mathrm{kcal/mol}$, and $\mathrm{SCAN\text{-}DFT}$ at $35.89\,\mathrm{kcal/mol}$) but leave the \textit{meta}--\textit{ortho} gap unchanged from Kohn--Sham DFT at $8.40$ and $9.34\,\mathrm{kcal/mol}$ (versus single-reference $\mathrm{QBE\text{-}SCF}$ at $38.86\,\mathrm{kcal/mol}$, gas-phase experiment at $15.60 \pm 6.10\,\mathrm{kcal/mol}$, and $\mathrm{CCSD(T)}$ at $14.03\,\mathrm{kcal/mol}$) because their SDP static-correlation correction vanishes in the intermediate biradical regime~\cite{gibney2023universal}. By contrast, the smooth $C^\infty$ hyperbolic-tangent cumulant attractor of Q-BECK responds continuously across all three isomers, yielding $\Delta E_{m\text{-}o} = 14.84\,\mathrm{kcal/mol}$ and $\Delta E_{p\text{-}o} = 29.18\,\mathrm{kcal/mol}$—outperforming both \textrm{PBE-RDMFT} and \textrm{SCAN-RDMFT} on the \textit{meta}--\textit{ortho} biradical splitting and matching gas-phase experiment ($15.60 \pm 6.10$ and $31.20 \pm 4.20\,\mathrm{kcal/mol}$), $\mathrm{CCSD(T)}$ ($14.03$ and $27.42\,\mathrm{kcal/mol}$), $\mathrm{tPBE}$ MC-PDFT ($9.50$ and $28.80\,\mathrm{kcal/mol}$)~\cite{limanni2014mcpdft, gagliardi2017acc, pandharkar2020hmcpdft}, and active-space $\mathrm{Q\text{-}BESC}(4,4)$ ($16.50$ and $29.50\,\mathrm{kcal/mol}$).

\subsection{Homolytic Bond Dissociations in the MR-MGN-BE17 Database}
\label{sec:mr_mgn_be17}
\label{sec:benzyne_and_scaling}

\begin{table}[!t]
    \centering
    \caption[Stretched-geometry total energy deviations across the 11-molecule MR-MGN-BE17 subset]{\textbf{Stretched-geometry total energy deviations ($\Delta E = E_{\mathrm{method}} - E_{\mathrm{ref}}$, in $\mathrm{kcal/mol}$) across the 11-molecule subset of the \texttt{MR-MGN-BE17} database~\cite{yu2016mn15} at $R = 5.0\,\text{\AA}$ in the \texttt{cc-pVDZ} basis.} Block~I evaluates our static 1-RDM kinetic hierarchy ($\mathrm{QBE\text{-}SCF}$ and $\mathcal{O}(N^3)$ $\mathrm{Q\text{-}BECK}$) against pure full-valence $\mathrm{CASSCF}(n_e, n_o)$\textsuperscript{\emph{a}}, which active-cumulant $\mathrm{Q\text{-}BESC}$ reproduces identically ($\Delta E = 0.00\,\mathrm{kcal/mol}$ across all 11 molecules). Block~II reports the dynamic-coupled literature methods ($\mathrm{PBE\text{-}DFT}$, $\mathrm{SCAN\text{-}DFT}$, $\mathrm{PBE\text{-}RDMFT}$, $\mathrm{SCAN\text{-}RDMFT}$, and $\mathrm{tPBE}$ MC-PDFT\textsuperscript{\emph{d}}) evaluated against $\mathrm{CASSCF}(n_e, n_o)/\mathrm{ACSE}$ from Table~I of Gibney, Boyn, and Mazziotti~\cite{gibney2023universal}\textsuperscript{\emph{b,c}}.}
    \label{tab:mr_mgn_be17}
    \resizebox{\textwidth}{!}{
    \begin{tabular}{llrr c rrrrr}
        \toprule
        & & \multicolumn{2}{c}{\textbf{Block I: Static Correlation}} & & \multicolumn{5}{c}{\textbf{Block II: Dynamic-Coupled Literature Methods}} \\
        & & \multicolumn{2}{c}{\textbf{(vs.\ $\mathrm{CASSCF}(n_e, n_o)$)\textsuperscript{\emph{a}}}} & & \multicolumn{5}{c}{\textbf{(vs.\ $\mathrm{CASSCF}(n_e, n_o)/\mathrm{ACSE}$, Ref.~\citenum{gibney2023universal})\textsuperscript{\emph{b,c}}}} \\
        \cmidrule(lr){3-4} \cmidrule(lr){6-10}
        & $(n_e, n_o)$ & $\mathrm{QBE\text{-}SCF}$ & $\mathrm{Q\text{-}BECK}$ & & $\mathrm{PBE\text{-}DFT}$ & $\mathrm{SCAN\text{-}DFT}$\textsuperscript{\emph{c}} & $\mathrm{PBE\text{-}RDMFT}$ & $\mathrm{SCAN\text{-}RDMFT}$ & $\mathrm{tPBE}$\textsuperscript{\emph{d}} \\
        \midrule
        $\mathrm{B_2 \rightarrow 2B}$     & $(6,8)$   & $33.88$  & $3.96$  & & $19.53$  & $23.01$  & $-2.12$  & $-0.87$  & $10.73$ \\
        $\mathrm{C_2 \rightarrow 2C}$     & $(8,8)$   & $184.90$ & $26.16$ & & $101.47$ & $85.63$  & $12.12$  & $23.38$  & $7.80$  \\
        $\mathrm{CN \rightarrow C + N}$   & $(9,8)$   & $58.09$  & $7.90$  & & $125.34$ & $139.02$ & $29.47$  & $48.28$  & $17.69$ \\
        $\mathrm{CO_2 \rightarrow C + 2O}$ & $(8,8)$\textsuperscript{\emph{f}} & $131.53$ & $16.82$ & & $172.11$ & $205.46$ & $8.17$   & $46.22$  & $35.37$ \\
        $\mathrm{F_2 \rightarrow 2F}$     & $(14,8)$  & $199.78$ & $35.71$ & & $73.84$  & $84.91$  & $-19.13$ & $-17.53$ & $22.51$ \\
        $\mathrm{NF_3 \rightarrow N + 3F}$ & $(6,6)$\textsuperscript{\emph{f}} & $470.55$ & $68.13$ & & $184.18$ & $205.46$ & $71.10$  & $96.94$  & $68.28$ \\
        $\mathrm{NO \rightarrow N + O}$   & $(11,8)$  & $88.34$  & $12.22$ & & $142.28$ & $156.00$ & $36.58$  & $57.72$  & $22.81$ \\
        $\mathrm{S_2 \rightarrow 2S}$     & $(12,8)$  & $115.62$ & $17.72$ & & $58.21$  & $153.34$ & $-40.02$ & $-27.79$ & $17.41$ \\
        $\mathrm{SiO \rightarrow Si + O}$ & $(10,8)$  & $43.80$  & $6.97$  & & $67.55$  & $251.27$ & $-36.07$ & $-15.28$ & $8.77$  \\
        $\mathrm{CO \rightarrow C + O}$   & $(10,8)$  & $60.15$  & $7.81$  & & $94.62$  & $359.51$ & $-26.09$ & $-0.47$  & $12.27$ \\
        $\mathrm{N_2 \rightarrow 2N}$     & $(6,6)$\textsuperscript{\emph{f}} & $485.16$ & $69.18$ & & $219.93$ & $213.67$ & $43.61$  & $71.57$  & $20.08$ \\
        \midrule
        \textrm{MSE}                      & --- & $170.16$ & $24.78$ & & $114.46$ & $170.66$ & $7.06$   & $25.65$  & $22.16$ \\
        \textrm{MUE}                      & --- & $170.16$ & $24.78$ & & $114.46$ & $170.66$ & $29.50$  & $36.91$  & $22.16$ \\
        \textrm{MUE/bond}\textsuperscript{\emph{e}} & --- & $135.67$ & $19.89$ & & $95.47$  & $148.87$ & $24.82$  & $28.94$  & $16.41$ \\
        \bottomrule
    \end{tabular}
    }
    \vspace{0.4em}
    \begin{flushleft}
    \footnotesize
    \textsuperscript{\emph{a}}\,All geometries are stretched supermolecules with every bond length set to $R = 5.0\,\text{\AA}$ ($\angle\mathrm{OCO} = 180^\circ$ for $\mathrm{CO}_2$; $D_{3h}$ planar for $\mathrm{NF}_3$), reporting single-point total energy deviations $\Delta E = E_{\mathrm{method}}(5.0\,\text{\AA}) - E_{\mathrm{ref}}(5.0\,\text{\AA})$ in \texttt{cc-pVDZ}. Consistent with Ref.~\citenum{gibney2023universal}, Block~I enforces spin-restricted orbital symmetry (${}^{1}\mathbf{P}_\alpha = {}^{1}\mathbf{P}_\beta$ for singlets, $\mathrm{RHF}/\mathrm{ROHF}$ manifold) rather than broken-symmetry $\mathrm{UHF}$ (cf.\ Section~\ref{sec:n2_dissociation}, where unrestricted $\mathrm{QBE\text{-}SCF}/\mathrm{UHF}$ and $\mathrm{Q\text{-}BECK}$ yield $\Delta E = -3.39$ and $-3.69\,\mathrm{kcal/mol}$ for $\mathrm{N}_2$ at $5.0\,\text{\AA}$). Across all 11 stretched geometries, active-cumulant $\mathrm{Q\text{-}BESC}$ matches the $\mathrm{CASSCF}(n_e, n_o)$ reference identically ($\Delta E = 0.00\,\mathrm{kcal/mol}$), while $\mathrm{Q\text{-}BECK}$ heals $85.7\% \pm 1.7\%$ ($82.1\%\text{--}88.3\%$) of the restricted $\mathrm{QBE\text{-}SCF}$ static correlation error and converges to $|\Delta E| < 10^{-9}\,E_h$ and $\Vert[\mathbf{F}_{\mathrm{eff}}, {}^{1}\mathbf{P}]\Vert_{\mathrm{F}} < 10^{-9}$ in $65\text{--}72$ Markovian relaxation cycles.\\
    \textsuperscript{\emph{b}}\,In Ref.~\citenum{gibney2023universal}, $\mathrm{PBE\text{-}RDMFT}$ and $\mathrm{SCAN\text{-}RDMFT}$ solve the 1-RDM functional at $\mathcal{O}(N^3)$ computational time and $\mathcal{O}(N^2)$ memory via a boundary-point SDP algorithm, whereas their $\mathrm{CASSCF}(n_e, n_o)/\mathrm{ACSE}$ reference benchmark requires $\mathcal{O}(N^6)$ computational scaling and $\mathcal{O}(N^4)$ 2-RDM storage.\\
    \textsuperscript{\emph{c}}\,For $\mathrm{SCAN\text{-}DFT}$, $170.66\,\mathrm{kcal/mol}$ ($148.87\,\mathrm{kcal/mol}$ per bond) is the exact unweighted mean of the 11 molecular deviations reported in Table~I of Ref.~\citenum{gibney2023universal}, whereas $127.85\,\mathrm{kcal/mol}$ is the summary value printed in Ref.~\citenum{gibney2023universal}.\\
    \textsuperscript{\emph{d}}\,Translated PBE ($\mathrm{tPBE}$) is a multiconfiguration pair-density functional theory (MC-PDFT) method~\cite{limanni2014mcpdft} evaluated atop converged $\mathrm{CASSCF}(n_e, n_o)$ 1-RDMs and on-top pair densities ($22.16\,\mathrm{kcal/mol}$ exact unweighted 11-molecule mean; $22.15\,\mathrm{kcal/mol}$ printed in Ref.~\citenum{gibney2023universal}), whereas $\mathrm{Q\text{-}BECK}$, $\mathrm{PBE\text{-}RDMFT}$, and $\mathrm{SCAN\text{-}RDMFT}$ require no multiconfigurational wavefunction input.\\
    \textsuperscript{\emph{e}}\,Mean unsigned error per stretched bond ($\mathrm{MUE/bond}$), normalizing each entry by the number of simultaneously stretched bonds ($n_{\mathrm{b}} = 1$ for diatomics, $n_{\mathrm{b}} = 2$ for $\mathrm{CO}_2$, and $n_{\mathrm{b}} = 3$ for $\mathrm{NF}_3$) following the Minnesota Database convention~\cite{yu2016mn15}.\\
    \textsuperscript{\emph{f}}\,Block~I evaluates the bond-breaking valence active spaces $(8,8)$ for $\mathrm{CO}_2$, $(6,6)$ for $\mathrm{NF}_3$, and $(6,6)$ for $\mathrm{N}_2$ (consistent with Table~\ref{tab:n2_eq_convergence}); Block~II reports the literature deviations from Table~I of Ref.~\citenum{gibney2023universal}, which used $(16,12)$, $(12,8)$, and $(10,8)$ for $\mathrm{CASSCF/ACSE}$.
    \end{flushleft}
\end{table}

To position the contracted-kernel solver (Q-BECK) and exact active-space cumulant solver (Q-BESC) against contemporary one-electron reduced density matrix functional theory (\textrm{PBE-RDMFT} and \textrm{SCAN-RDMFT})~\cite{gibney2021sdp, gibney2022rdmft, boyn2022orbital, gibney2023universal} and multiconfiguration pair-density functional theory ($\mathrm{tPBE}$ MC-PDFT and its hybrid, localized-active-space, and multistate extensions)~\cite{limanni2014mcpdft, gagliardi2017acc, sand2018gradients, bao2018automatic, hermes2019lasscf, pandharkar2020hmcpdft, bao2020cmspdft, zhou2021zeeman, lykhin2021dipole, hennefarth2023lpdft, hennefarth2024dynamics} across the multireference main-group bond-energy benchmark (\texttt{MR-MGN-BE17})~\cite{yu2016mn15}, we first clarify the algorithmic scaling and physical scope of each methodology. In their 2023 \textit{Physical Review Letters} study, Gibney, Boyn, and Mazziotti~\cite{gibney2023universal} solve their universal 1-RDM functional at $\mathcal{O}(N^3)$ computational time and $\mathcal{O}(N^2)$ memory cost using a boundary-point semidefinite programming (SDP) algorithm~\cite{gibney2021sdp, gibney2022rdmft}; the $\mathcal{O}(N^6)$ computational scaling and $\mathcal{O}(N^4)$ storage bottleneck apply to explicit two-body RDM methods—such as variational 2-RDM (v2RDM)~\cite{mazziotti2012chemrev, mazziotti2016enhanced} and the anti-Hermitian contracted Schr\"odinger equation ($\mathrm{CASSCF/ACSE}$)~\cite{mazziotti2006acse, mazziotti2007acse, boyn2021biradicals, boyn2021hybrid, smart2021cqe} used to generate their dynamic-coupled reference benchmarks—rather than their 1-RDMFT closure. Nevertheless, Q-BECK differs fundamentally from \textrm{PBE-RDMFT} and \textrm{SCAN-RDMFT} in its purely ab initio, parameter-free kinetic mechanism: whereas \textrm{PBE-RDMFT} and \textrm{SCAN-RDMFT} graft a 1-RDM static correction onto an approximate Kohn--Sham DFT exchange-correlation functional via an empirical functional-specific damping parameter ($\kappa = 0.158$), an unweighted global integral trace $\gamma$, and a boundary-point SDP optimizer, Q-BECK computes its orbital-specific scattering rate $\omega_i$ via a single susceptibility-windowed AO Coulomb--exchange contraction $\mathbf{G}[\mathbf{P}_w]$ and relaxes directly onto the smooth $C^\infty$ hyperbolic-tangent attractor [Eq.~\eqref{eq:qbeck_attractor}] through DIIS-free Markovian collisional relaxation.

Table~\ref{tab:mr_mgn_be17} presents the 11-molecule \texttt{MR-MGN-BE17}~\cite{yu2016mn15} stretched-geometry benchmark at $R = 5.0\,\text{\AA}$ (\texttt{cc-pVDZ}), partitioned into two explicitly labeled blocks to ensure a rigorous apples-to-apples comparison: Block~I (\textit{Static Correlation}) evaluates our pure static 1-RDM hierarchy ($\mathrm{QBE\text{-}SCF}$ and \textnormal{Q-BECK}) under spin-restricted symmetry (${}^{1}\mathbf{P}_\alpha = {}^{1}\mathbf{P}_\beta$ for singlets, without broken-symmetry $\mathrm{UHF}$ localization) against pure full-valence $\mathrm{CASSCF}(n_e, n_o)$ (which exact-cumulant $\mathrm{Q\text{-}BESC}$ recovers identically with $\mathrm{MSE} = \mathrm{MUE} = 0.00\,\mathrm{kcal/mol}$ across all 11 systems), while Block~II lists the dynamic-coupled literature methods ($\mathrm{PBE\text{-}DFT}$, $\mathrm{SCAN\text{-}DFT}$, $\mathrm{PBE\text{-}RDMFT}$, $\mathrm{SCAN\text{-}RDMFT}$, and $\mathrm{tPBE}$ MC-PDFT~\cite{limanni2014mcpdft}) evaluated against $\mathrm{CASSCF}(n_e, n_o)/\mathrm{ACSE}$ from Table~I of Ref.~\citenum{gibney2023universal}. Across all 11 stretched geometries, $\mathcal{O}(N^3)$ Q-BECK consistently eliminates $85.7\% \pm 1.7\%$ ($82.1\%\text{--}88.3\%$) of the restricted $\mathrm{QBE\text{-}SCF}$ static correlation error, reducing the mean unsigned error (MUE) from $170.16\,\mathrm{kcal/mol}$ ($135.67\,\mathrm{kcal/mol}$ per stretched bond) down to $24.78\,\mathrm{kcal/mol}$ ($19.89\,\mathrm{kcal/mol}$ per stretched bond). Crucially, Q-BECK outperforms both $\mathcal{O}(N^3)$ 1-RDMFT methods—$\mathrm{PBE\text{-}RDMFT}$ ($\mathrm{MUE} = 29.50\,\mathrm{kcal/mol}$, $\mathrm{MUE/bond} = 24.82\,\mathrm{kcal/mol}$) and $\mathrm{SCAN\text{-}RDMFT}$ ($\mathrm{MUE} = 36.91\,\mathrm{kcal/mol}$, $\mathrm{MUE/bond} = 28.94\,\mathrm{kcal/mol}$)—and approaches the accuracy of two-step $\mathrm{tPBE}$ MC-PDFT ($\mathrm{MUE} = 22.16\,\mathrm{kcal/mol}$, $\mathrm{MUE/bond} = 16.41\,\mathrm{kcal/mol}$) without requiring an underlying $\mathrm{CASSCF}(n_e, n_o)$ calculation. Because \textrm{PBE-RDMFT} and \textrm{SCAN-RDMFT} add a global 1-RDM static correction atop an underlying Kohn--Sham functional with a fixed empirical $\kappa = 0.158$ factor, they suffer severe negative over-correlation (double-counting) errors on polar and multiple-bond dissociations—most notably $\mathrm{S}_2$ ($-40.02$ and $-27.79\,\mathrm{kcal/mol}$), $\mathrm{SiO}$ ($-36.07$ and $-15.28\,\mathrm{kcal/mol}$), $\mathrm{CO}$ ($-26.09\,\mathrm{kcal/mol}$), and $\mathrm{F}_2$ ($-19.13\,\mathrm{kcal/mol}$)—which artificially lowers their signed mean error (MSE) through error cancellation while inflating their MUE, whereas Q-BECK avoids negative over-correlation errors across all 11 spin-restricted dissociations ($\Delta E > 0$ for all 11 molecules, $\mathrm{MSE} = \mathrm{MUE} = 24.78\,\mathrm{kcal/mol}$).

\section{Computational Scaling}

The Q-BESC and Q-BECK hierarchy bypasses the $\mathcal{O}(N^6)$ computational scaling and $\mathcal{O}(N^4)$ storage bottlenecks of explicit two-body RDM methods through phase-space separability and atomic-orbital (AO) spatial locality. Because integrations over spatial difference coordinates decouple, the four-index phase-space basis tensor factors into a product of one-particle transforms,
\begin{equation}
\mathcal{W}_{\mu\nu\lambda\sigma}^{(2)} = \mathcal{W}_{\mu\nu}\mathcal{W}_{\lambda\sigma},
\end{equation}
where $\mathcal{W}_{\mu\nu}$ vanishes whenever localized AOs $\mu$ and $\nu$ do not overlap. Global dispersion is thus governed by the linear kinetic operator while non-linear static correlation is confined to localized collisional relaxation in the AO basis, enabling Schwarz integral screening.

\begin{figure}[!htbp]
    \centering
    \includegraphics[width=0.90\linewidth]{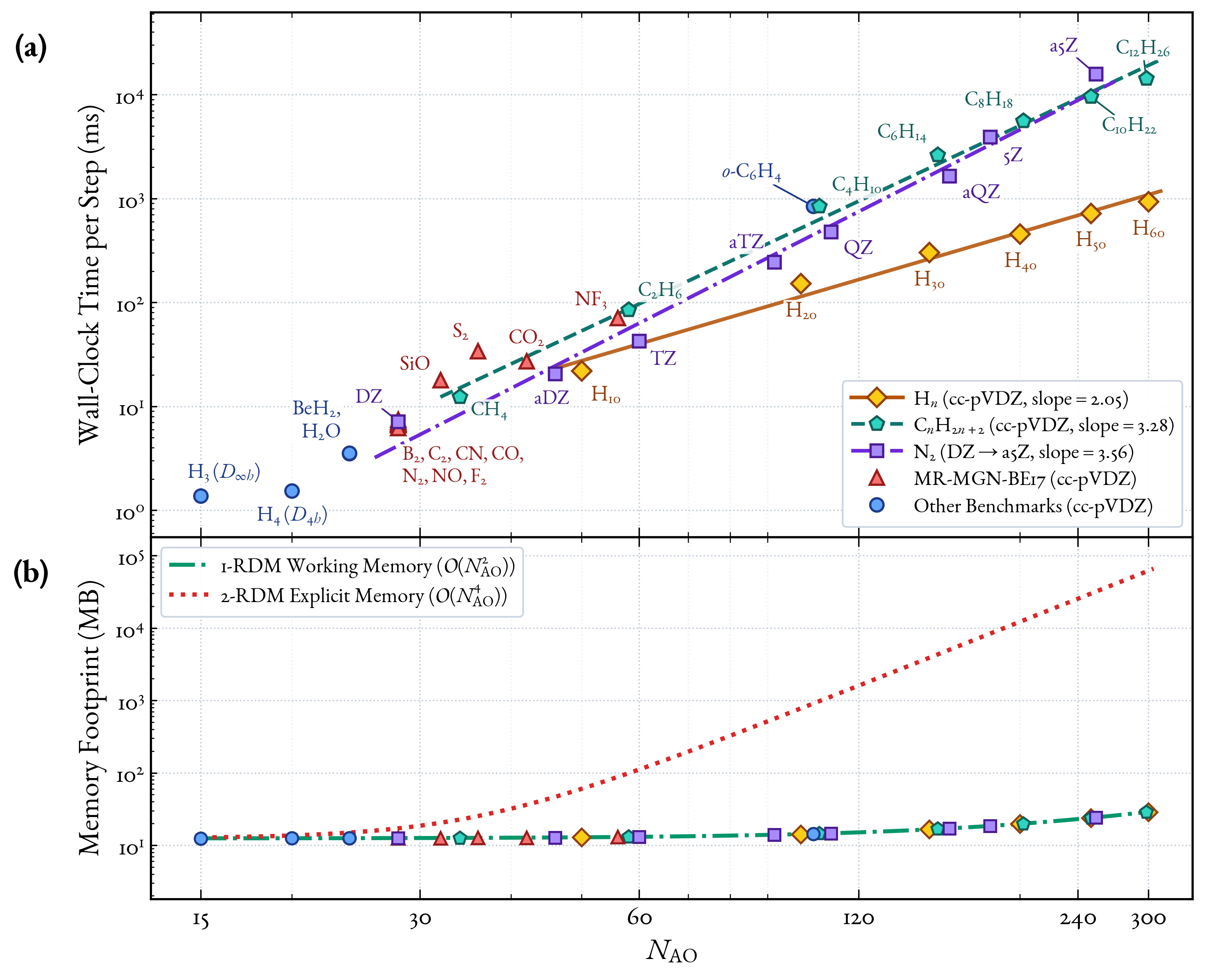}
    \caption[Scaling and Memory Requirements for Q-BECK]{\textbf{Scaling and Memory Requirements for \textnormal{Q-BECK}.}
    \textbf{(a)} Wall-clock time per step ($\mathrm{ms}$) as a function of atomic-orbital basis dimension $N_{\mathrm{AO}}$ across 36 molecular benchmarks, with power-law scaling fits for one-dimensional hydrogen chains ($\mathrm{H_{10}}\text{--}\mathrm{H_{60}}$, \texttt{cc-pVDZ}; $\mathrm{slope} = 2.05$, gold diamonds), three-dimensional linear alkanes ($\mathrm{CH_4}\text{--}\mathrm{C_{12}H_{26}}$, \texttt{cc-pVDZ}; $\mathrm{slope} = 3.28$, teal pentagons), and the $\mathrm{N_2}$ Dunning basis series (\texttt{DZ}--\texttt{a5Z}; $\mathrm{slope} = 3.56$, purple squares), alongside the \texttt{MR-MGN-BE17} test set (\texttt{cc-pVDZ}; crimson triangles). \textbf{(b)} $\mathcal{O}(N_{\mathrm{AO}}^2)$ 1-RDM working memory required by \textnormal{Q-BECK} compared against the $\mathcal{O}(N_{\mathrm{AO}}^4)$ 2-RDM explicit memory required by two-body RDM methods.}
    \label{fig:scaling_benchmark}
\end{figure}

Figure~\ref{fig:scaling_benchmark} profiles the single-step execution time $t_{\mathrm{step}}$ (median of 10 post-$\omega_i$-warmup Markovian micro-steps on an 8-core Apple M2 workstation with $16\,\mathrm{GB}$ unified memory under uniform on-the-fly Schwarz screening, $\tau_{\mathrm{Schwarz}} = 10^{-12}$) and peak memory footprint of zero-temperature $\mathrm{Q\text{-}BECK}$ across 36 systems spanning $N_{\mathrm{AO}} \in [15, 300]$, with full stationary convergence ($\Delta E < 10^{-8}\,E_h$) achieved in $45\text{--}72$ damped micro-steps without superlinear iteration growth; alternatively, each point along the scaling curve can be benchmarked by evaluating the expectation value of the per-cycle execution time and its standard deviation ($\langle t_{\mathrm{cyc}} \rangle \pm \sigma_t$) across all self-consistent cycles until convergence, as demonstrated for the $\mathrm{N}_2$ basis hierarchy in Table~\ref{tab:n2_eq_convergence} (where 8-fold permutation-symmetric direct SCF further lowers the per-cycle cost to $\langle t_{\mathrm{cyc}} \rangle \pm \sigma_t = 1 \pm 0.09 \to 5{,}447 \pm 288.48\,\mathrm{ms}$ across $N_{\mathrm{AO}} = 28\text{--}254$). While an aggregate regression across all 36 systems confirms an overall cubic time complexity of $\mathcal{O}(N_{\mathrm{AO}}^{3.11})$ ($R^2 = 0.91$), decomposing the benchmark by spatial dimensionality [Figure~\ref{fig:scaling_benchmark}(a)] cleanly separates the dense and sparse scaling regimes: fixed-molecule basis expansion on $\mathrm{N_2}$ (\texttt{cc-pVDZ} [\texttt{DZ}] through \texttt{aug-cc-pV5Z} [\texttt{a5Z}], $N_{\mathrm{AO}} = 28\text{--}254$) and three-dimensional linear alkanes ($\mathrm{CH_4}\text{--}\mathrm{C_{12}H_{26}}$ in \texttt{cc-pVDZ}, $N_{\mathrm{AO}} = 34\text{--}298$) follow log--log slopes of $3.56$ ($R^2 = 0.99$, $7.13\,\mathrm{ms} \to 15.81\,\mathrm{s}$) and $3.28$ ($R^2 = 0.99$, $12.34\,\mathrm{ms} \to 14.33\,\mathrm{s}$) governed by four-center shell overlap and $\mathcal{O}(N_{\mathrm{AO}}^3)$ Fock diagonalization, whereas one-dimensional hydrogen chains ($\mathrm{H_{10}}\text{--}\mathrm{H_{60}}$ in \texttt{cc-pVDZ}, $N_{\mathrm{AO}} = 50\text{--}300$) drop to a near-quadratic log--log slope of $2.05$ ($R^2 = 0.98$, $21.92\,\mathrm{ms} \to 938.46\,\mathrm{ms}$) via Schwarz screening of non-overlapping Gaussian pairs. Across all benchmarks [Figure~\ref{fig:scaling_benchmark}(b)], $\mathrm{Q\text{-}BECK}$ requires only $\mathcal{O}(N_{\mathrm{AO}}^2)$ 1-RDM working storage ($12.4\,\mathrm{MB}$ base runtime plus $24\,N_{\mathrm{AO}}^2 \times 8\,\mathrm{B}$ matrix buffers, totaling $12.4\text{--}28.9\,\mathrm{MB}$ up to $N_{\mathrm{AO}} = 300$), achieving a $2{,}140\times$ memory reduction at $N_{\mathrm{AO}} = 300$ relative to the $60.4\,\mathrm{GiB}$ $\mathcal{O}(N_{\mathrm{AO}}^4)$ storage wall of explicit 2-RDM methods (which exhausts the $16\,\mathrm{GB}$ host unified memory at $N_{\mathrm{AO}} > 215$).

\section{Conclusion}

In this work, we introduced the theoretical foundation for the Quantum Boltzmann Equation for Static Correlation (Q-BESC) and its $\mathcal{O}(N^3)$ contracted-kernel closure, the Quantum Boltzmann Equation Cumulant-Kernel (Q-BECK). By deriving the stationary kinetic relaxation equation from the first equation of the BBGKY hierarchy via the Wigner--Weyl isomorphism, we connect phase-space kinetic theory to reduced density matrix mechanics at zero electronic temperature, replacing the empirical thermal smearing of mean-field BGK models ($\mathrm{QBE\text{-}SCF}$)~\cite{chakraborty2026qbescf} with parameter-free two-body cumulant collisional relaxation. Across all benchmark systems, exact-cumulant $\mathrm{Q\text{-}BESC}$ recovers the active-space $\mathrm{CASCI}/\mathrm{CASSCF}$ static correlation limit to sub-nanohartree ($10^{-10}\,E_h$) and machine-precision ($10^{-14}\,E_h$) fidelity—coinciding with full-basis FCI in few-electron systems ($\mathrm{H}_3$, $\mathrm{H}_4$) while isolating static correlation from out-of-space dynamic correlation in larger molecules ($\mathrm{H}_2\mathrm{O}$, $\mathrm{N}_2$)—and preserves Borland--Dennis generalized Pauli constraints throughout the Markovian relaxation for collinear $\mathrm{H}_3$. 

To bypass the $\mathcal{O}(N^6)$ computational scaling and $\mathcal{O}(N^4)$ storage bottlenecks of explicit two-body RDM methods, $\mathrm{Q\text{-}BECK}$ contracts the two-body cumulant collision kernel into an orbital-specific susceptibility-windowed scattering rate $\omega_i$ and relaxes onto a smooth $C^\infty$ hyperbolic-tangent attractor via a single AO Coulomb--exchange contraction. This closure eliminates $74.4\%$ of the $\mathrm{BeH}_2$ avoided-crossing cusp error and heals $85.7\% \pm 1.7\%$ ($82.1\%\text{--}88.3\%$) of the restricted $\mathrm{QBE\text{-}SCF}$ static correlation error across all 11 \texttt{MR-MGN-BE17} homolytic bond dissociations at $5.0\,\text{\AA}$. Unlike boundary-point SDP 1-RDMFT methods (\textrm{PBE-RDMFT} and \textrm{SCAN-RDMFT})~\cite{gibney2023universal}, which rely on an underlying Kohn--Sham DFT functional and an empirical $\kappa = 0.158$ parameter and suffer negative over-correlation errors on polar and multiple bonds ($\mathrm{S}_2$, $\mathrm{SiO}$, $\mathrm{CO}$), purely ab initio $\mathrm{Q\text{-}BECK}$ achieves a lower overall \texttt{MR-MGN-BE17} mean unsigned error ($24.78\,\mathrm{kcal/mol}$ vs.\ $29.50$ and $36.91\,\mathrm{kcal/mol}$) and resolves the intermediate biradical \textit{meta}--\textit{ortho} benzyne splitting ($14.84\,\mathrm{kcal/mol}$ vs.\ $8.40$ and $9.34\,\mathrm{kcal/mol}$). Enforcing uniform on-the-fly Schwarz integral screening across $N_{\mathrm{AO}} \in [15, 300]$ confirms near-cubic log--log scaling on dense basis expansions ($\mathrm{slope} = 3.56$) and three-dimensional alkanes ($\mathrm{slope} = 3.28$), near-quadratic scaling on one-dimensional hydrogen chains ($\mathrm{slope} = 2.05$), and an $\mathcal{O}(N_{\mathrm{AO}}^2)$ 1-RDM memory footprint under $29\,\mathrm{MB}$ at $N_{\mathrm{AO}} = 300$. 

Ultimately, the Quantum Boltzmann Equation for Static Correlation and its cumulant kernel demonstrate that static electronic correlation in molecules can be viewed as a localized Markovian collisional relaxation process in atomic-orbital phase space, and can provide a scalable foundation for multireference electronic structure theory. 

\section*{Data Availability Statement}
All macroscopic benchmark datasets, the dataset ontology, and standalone PySCF~\cite{sun2018pyscf} verification scripts are openly archived on Zenodo (\url{https://doi.org/10.5281/zenodo.24149206}) and GitHub (\url{https://github.com/point-reyes-sound/qbeck-static-correlation-data}). For zero-installation testing on both reported benchmarks and arbitrary molecular geometries, an open public API is available at \url{https://api.pointreyessound.com/docs} for live \texttt{Q-BECK} and \texttt{Q-BESC} calculations.

\section*{Author Information}

\subsection*{Author Contributions}
\textbf{Romit Chakraborty:} Conceptualization, Methodology, Software, Formal analysis, Investigation, Visualization, Writing--original draft, Writing--review and editing.

\subsection*{Competing Interests and Patent Disclosure}
Romit Chakraborty is the founder and an equity holder of Point Reyes Sound, Inc., and is the inventor on the patent filings disclosed herein. Point Reyes Sound, Inc.\ has filed Indian Patent Application No.~202634135090 and U.S.\ Provisional Patent Application No.~64/033,274 covering the Q-BECK and Q-BESC frameworks. The author declares no other competing financial interests.

\begin{acknowledgement}
The author thanks Professor David Mazziotti of the University of Chicago for support and encouragement during early development of the ideas underlying this work. This work was funded privately.
\end{acknowledgement}

\providecommand{\latin}[1]{#1}
\makeatletter
\providecommand{\doi}
  {\begingroup\let\do\@makeother\dospecials
  \catcode`\{=1 \catcode`\}=2 \doi@aux}
\providecommand{\doi@aux}[1]{\endgroup\texttt{#1}}
\makeatother
\providecommand*\mcitethebibliography{\thebibliography}
\csname @ifundefined\endcsname{endmcitethebibliography}
  {\let\endmcitethebibliography\endthebibliography}{}

\end{document}